%% file: paper.tex
\documentclass[submission,letterpaper,Phys]{SciPost}
\pdfoutput=1
\usepackage{amsmath,amssymb,mathtools,xspace}
\usepackage{booktabs,multirow,graphicx,tabularx,slashed}
\usepackage{hyperref}
\usepackage{color,xcolor}
\usepackage[normalem]{ulem}
\usepackage{enumitem}
\usepackage{feynmp}
\usepackage{braket}
\usepackage{tikz}
\usetikzlibrary{arrows}
\usetikzlibrary{shapes.geometric, arrows}

\makeatletter
\@ifundefined{pdfoutput}{}{\DeclareGraphicsRule{*}{mps}{*}{}}
\makeatother

\makeatletter
\DeclareRobustCommand*{\bfseries}{%
   \not@math@alphabet\bfseries\mathbf
   \fontseries\bfdefault\selectfont
   \boldmath
}
\makeatother

\input{incl_declare}

\begin{document}
\begin{fmffile}{feynman}

\begin{center}{\Large \textbf{
How to GAN away Detector Effects
}}\end{center}

\begin{center}
Marco Bellagente\textsuperscript{1},
Anja Butter\textsuperscript{1},
Gregor Kasieczka\textsuperscript{2},
Tilman Plehn\textsuperscript{1}, and
Ramon Winterhalder\textsuperscript{1}
\end{center}

\begin{center}
{\bf 1} Institut f\"ur Theoretische Physik, Universit\"at Heidelberg, Germany\\
{\bf 2} Institut f\"ur Experimentalphysik, Universit\"at Hamburg, Germany
bellagente@thphys.uni-heidelberg.de
\end{center}

\begin{center}
\today
\end{center}


\tikzstyle{int}=[thick,draw, minimum size=2em]

\section*{Abstract}
{\bf LHC analyses directly comparing data and simulated events bear the
  danger of using first-principle predictions only as a black-box
  part of event simulation. We show how simulations, for instance, of
  detector effects can instead be inverted using generative
  networks. This allows us to reconstruct parton level information
  from measured events. Our results illustrate how, in general, fully
  conditional generative networks can statistically invert Monte Carlo
  simulations. As a technical by-product we show how a maximum mean
  discrepancy loss can be staggered or cooled.}

\vspace{10pt}
\noindent\rule{\textwidth}{1pt}
\tableofcontents\thispagestyle{fancy}
\noindent\rule{\textwidth}{1pt}
\vspace{10pt}

\newpage
\section{Introduction}
\label{sec:intro}

Our understanding of LHC data from first principles is a unique
strength of particle physics. It is based on a simulation chain which
starts from a hard process described by perturbative QCD, and then
adds the logarithmically enhanced QCD parton shower, fragmentation,
hadronization, and finally a fast or complete detector
simulation~\cite{black_book}. This simulation chain is publicly
available and relies on extremely efficient, fast, and reliable Monte
Carlo techniques.

Unfortunately, there is a price for this efficiency: while in
principle such a Monte Carlo simulation as a Markov process can be
inverted at least statistically, in practice we have to employ
approximations. This asymmetry has serious repercussions for LHC
analyses, where for instance we do not have access to the likelihood
ratio of the hard process. Even worse, it seriously limits our
interpretation of LHC results because we cannot easily show results in
terms of observables accessible by perturbative QCD. For typical ATLAS
or CMS limit reporting this might seem less relevant, but every so
often we want to be able to understand such a result more
quantitatively.

\begin{figure}[b!]
\begin{center}
\input{incl_feynman}
\end{center}
\caption{Sample Feynman diagram contributing to $WZ$ production, with
  intermediate on-shell particles labelled.}
\label{fig:feyn_intro}
\end{figure}
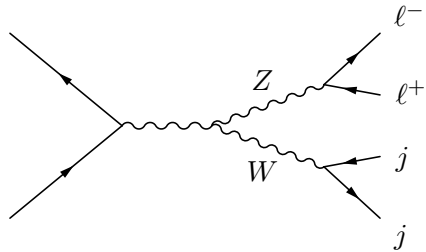

We propose to use generative networks or GANs~\cite{goodfellow} to
invert Monte Carlo simulations. There are many examples showing that
we can GAN such simulations, including phase space
integration~\cite{maxim,bendavid}, event
generation~\cite{dutch,gan_datasets,DijetGAN2, gan_phasespace},
detector
simulations~\cite{calogan1,calogan2,fast_accurate,aachen_wgan1,aachen_wgan2,ATLASShowerGAN,ATLASsimGAN},
unfolding~\cite{Datta:2018mwd}, and parton
showers~\cite{shower,locationGAN,monkshower,juniprshower}. The
question is if and how we can invert them.  We start with a naive GAN
inversion and see how a mismatch between local structures at parton
level and detector level leads to problems.  We then introduce the
first fully conditional GAN~\cite{cond_gan} (FCGAN) in particle
physics to invert a fast detector simulation~\cite{delphes} for the
process
\begin{align}
pp
\to ZW^\pm
\to (\ell^- \ell^+) \; (j j ) \; ,
\end{align}
as illustrated in Fig.~\ref{fig:feyn_intro}.  We will see how the
fully conditional setup gives us all the required properties of an
inverted detector simulation. 

We note that our approach is not targeted at combining detector
unfolding~\cite{Cowan:2002in,Blobel:2011fih,Balasubramanian:2019itp}
with optimized inference~\cite{madminer,madminer_tool,omnifold}.  A
powerful application for unfolded kinematic distributions to the hard
process could be global analyses. For instance in the electroweak and
Higgs sector exotics resonance searches turn out to be among the most
interesting input and pose a challenge when including
them~\cite{Biekotter:2018rhp}. In contrast, global analyses in the top
sector~\cite{Brivio:2019ius,Hartland:2019bjb,Buckley:2015lku}
successfully rely on unfolded information to different levels of the
hard process, for instance the top pair production
process~\cite{Khachatryan:2015oqa,Aad:2015mbv}. At the same time,
alternative methods like simplified template cross sections lose a
sizeable amount of information~\cite{Brehmer:2019gmn}.  The same
method would also allow us to directly compare first-principles QCD
predictions with modern LHC measurements.  In addition, our fast
inversion might help with advanced statistical techniques like the
matrix element
method~\cite{Kondo:1988yd,Martini:2015fsa,Gritsan:2016hjl,Martini:2017ydu,Kraus:2019qoq,Prestel:2019neg}.

But most importantly, our FCGAN first serves as an example how we can
invert Monte Carlo simulations to understand the physics behind modern
LHC analyses based on a direct comparison of data and
simulations. Here the GAN benefits from the excellent interpolations
properties of neural networks. Second, faithfully preserves local
structures leading to a large degree of model independence in the
unfolding procedure.

\section{GAN unfolding}
\label{sec:gan}

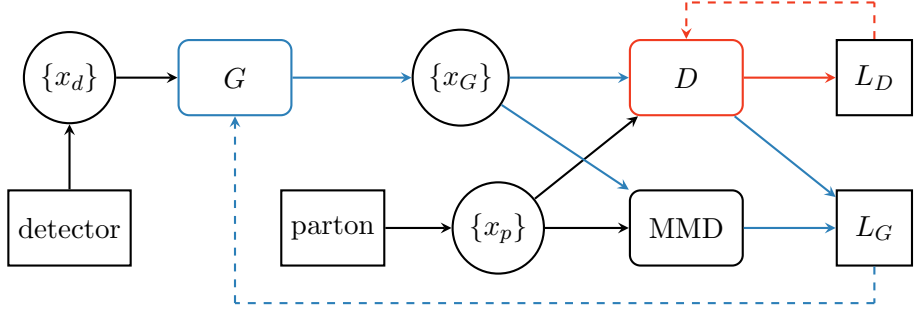
\begin{figure}[t]
\centering
\input{incl_network_gan}
\caption{Structure of a naive unfolding GAN. The input $\{ x_d \}$
  describes a batch of events sampled at detector level and $\{
  x_{G,p} \}$ denotes events sampled from the generator or
  parton-level data. The blue (red) arrows indicate which
  connections are used in the training of the generator
  (discriminator).}
\label{fig:GANs}
\end{figure}

A standard method for fast detector simulation is
smearing the outgoing particle momenta with a detector response
function. This allows us to generate and sample from a probability
distribution of smeared final-state momenta for a single parton-level
event. For the inversion we need to rely on event samples, as we can
see from a simple example: we start from a sharp $Z$-peak at the
parton level and broaden it with detector effects. Now we look at a
detector-level event in the tail and invert the detector simulations,
for which we need to know in which direction in the invariant mass the
preferred value $m_Z$ lies. This implies that unfolding detector
effects requires a model hypothesis, which can be thought of as a
condition in a probability of the inversion from the detector
level. The problem with this point of view is that the parton-level
distribution of the invariant mass requires a dynamic reconstruction
of the Breit-Wigner peak, which is not easily combined with a Markov
process. In any case, from this argument it is clear that unfolding
only makes sense at the level of large enough event samples.

For our example we rely on two event samples: we start with events at
the parton level, simulated with \madgraph~\cite{madgraph}. For
the second sample we first apply \pythia not including initial state radiation. Technically this
means that we only have to deal with a fixed number of partons in the
final state and that we can more easily match partons and jets. Further, we apply \delphes~\cite{delphes} as a fast detector
simulation and reconstruct the smeared jet 4-momenta with a jet
algorithm included in \fastjet~\cite{FastJet}. For lepton 4-momenta we can directly compare the
parton-level output with the detector-level output. From
Ref.~\cite{gan_phasespace} we know how to set up a GAN to either
generate detector-level events from parton-level events or vice
versa. In our current setup the events are unweighted set of four
4-vectors, two jets and two leptons, but it can be easily adapted to
weighted events. The final-state masses are fixed to the parton-level
values. Our hadronic final state is defined at the level of jet
4-vectors. This does not mean that in a possible application we take a
parton shower at face value. All we do is assume that there is a
correspondence between a hard parton and its hadronic final state, and
that the parton 4-momentum can be reconstructed with the help of a jet
algorithm. The question if for instance an anti-$k_T$ algorithm is an
appropriate description of sub-jet physics does not arise as long as
the jet algorithm reproduces the hard parton momentum.

Our GAN comprises a generator network $G$ competing against a
discriminator network $D$ in a min-max game, as illustrated in
Fig.~\ref{fig:GANs}. For the implementation we use \keras~(v2.2.4) \cite{keras} with a \tensorflow~(v1.14) backend \cite{tensorflow}.
As the starting point, $G$ is randomly
initialized to produce an output, typically with the same
dimensionality as the target space. It induces a probability
distribution $P_G(x)$ of a target space element $x$, in our case a
parton-level event. To be precise, the generator obtains a batch of
detector level event as input and generates a batch of parton level
events as output, \ie $G(\{x_d\}) = \{x_G\}$.  The discriminator is
given batches $\left\{x_G \right\}$ and $\left\{x_p \right\}$ sampled
from $P_G$ and the parton-level target distribution $P_p$. It is
trained as a binary classifier, such that $D\left(x \in
\left\{x_p\right\} \right) = 1$ and $D\left(x\right) = 0$
otherwise. Following the conventions of Ref.~\cite{gan_phasespace} the
discriminator loss function is defined as
\begin{align}
L_D = \left\langle - \log D\left(x\right) \right\rangle_{x \sim P_p} + \left\langle - \log \left(1-D\left(x\right)\right) \right\rangle_{x \sim P_G} \; .
\label{eq:D_loss}
\end{align}
We add a regularization and obtain the regularized Jensen-Shannon GAN
loss function~\cite{gan_stabilize_training}
\begin{align}
L_D^\text{(reg)} =
L_D
+ \lambda_D\,
\Langle \left(1- D(x)\right)^2 \vert \nabla \phi \vert^2 \Rangle_{x \sim P_p} 
+ \lambda_D\,
\Langle D(x)^2\, \vert \nabla \phi \vert^2 \Rangle_{x \sim P_G} \; ,
\label{eq:D_loss2}
\end{align}
with a properly chosen pre-factor $\lambda_D$ and where we define 
$\phi(x) = \log \frac{D(x)}{1-D(x)}$.
The discriminator training at fixed $P_p$ and $P_G$ alternates
with the generator training, which is trained to maximize the second
term in Eq.\eqref{eq:D_loss} using the truth encoded in
$D$. This is efficiently encoded in minimizing
\begin{align}
L_G = \left\langle - \log D\left(x\right) \right\rangle_{x \sim P_G} \; .
\label{eq:G_loss}
\end{align}
If the training of the generator and the discriminator with their
respective losses Eq.\eqref{eq:D_loss2} and Eq.\eqref{eq:G_loss} is
properly balanced, the distribution $P_G$ converges to the parton-level
distribution $P_p$, while the optimized discriminator is unable to
distinguish between real and generated samples.

If we want to describe phase space features, for instance at the LHC,
it is useful to add a maximum mean discrepancy (MMD)~\cite{mmd}
contribution to the loss function~\footnote{For all details on
  combining GANs with MMD we refer to the original
  paper~\cite{gan_phasespace}.}. It allows us to compare pre-defined
distributions, for instance the one-dimensional invariant mass of an
intermediate particle. Given batches of true and generated
parton-level events we define the additional contribution to the
generator loss as
\begin{align}
\text{MMD} =
\left[ \langle k\left(x,x'\right)\rangle_{x,x' \sim P_G} 
     + \langle k\left(y,y'\right)\rangle_{y,y' \sim P_p} 
     - 2 \langle k\left(x,y \right)\rangle_{x\sim P_G,y \sim P_p} \right]^{1/2} \; ,
\label{eq:MMD}
\end{align}
with another pre-factor $\lambda_G$. Note that we use $\text{MMD}$ instead
of $\text{MMD}^2$ to enhance the sensitivity of the model \cite{GMMN}. In
Ref.~\cite{gan_phasespace} we have compared common choices, like
Gaussian or Breit-Wigner kernels with a given width $\sigma$,
\begin{align}
k_\text{Gauss} \left(x,y\right) = \exp \frac{- \left(x - y\right)^2}{2 \sigma^2}
\qquad \text{or} \qquad 
k_\text{BW}\left(x,y\right) = \frac{\sigma^2}{\left(x - y\right)^2 + \sigma^2} \; .
\label{eq:kernels}
\end{align}
As a naive approach to GAN unfolding we use detector-level event
samples as generator input. The network input is always a set of four
4-vectors, one for each particle in the final state, with their masses
fixed~\cite{gan_phasespace}.  In the GAN setup we train our network to
map detector-level events to parton-level events.  Both networks
consist of 12 layers with 512 units per layer. With $\lambda_G=1$,
$\lambda_D=10^{-3}$ and a batch size of 512 events, we run for 1200
epochs and 500 iterations per epoch.\medskip

\begin{figure}[t]
\centering
\includegraphics[page = 2, width=0.49\textwidth]{figures/GAN_ratio}
\includegraphics[page = 3, width=0.49\textwidth]{figures/GAN_ratio} \\
\includegraphics[page = 1, width=0.49\textwidth]{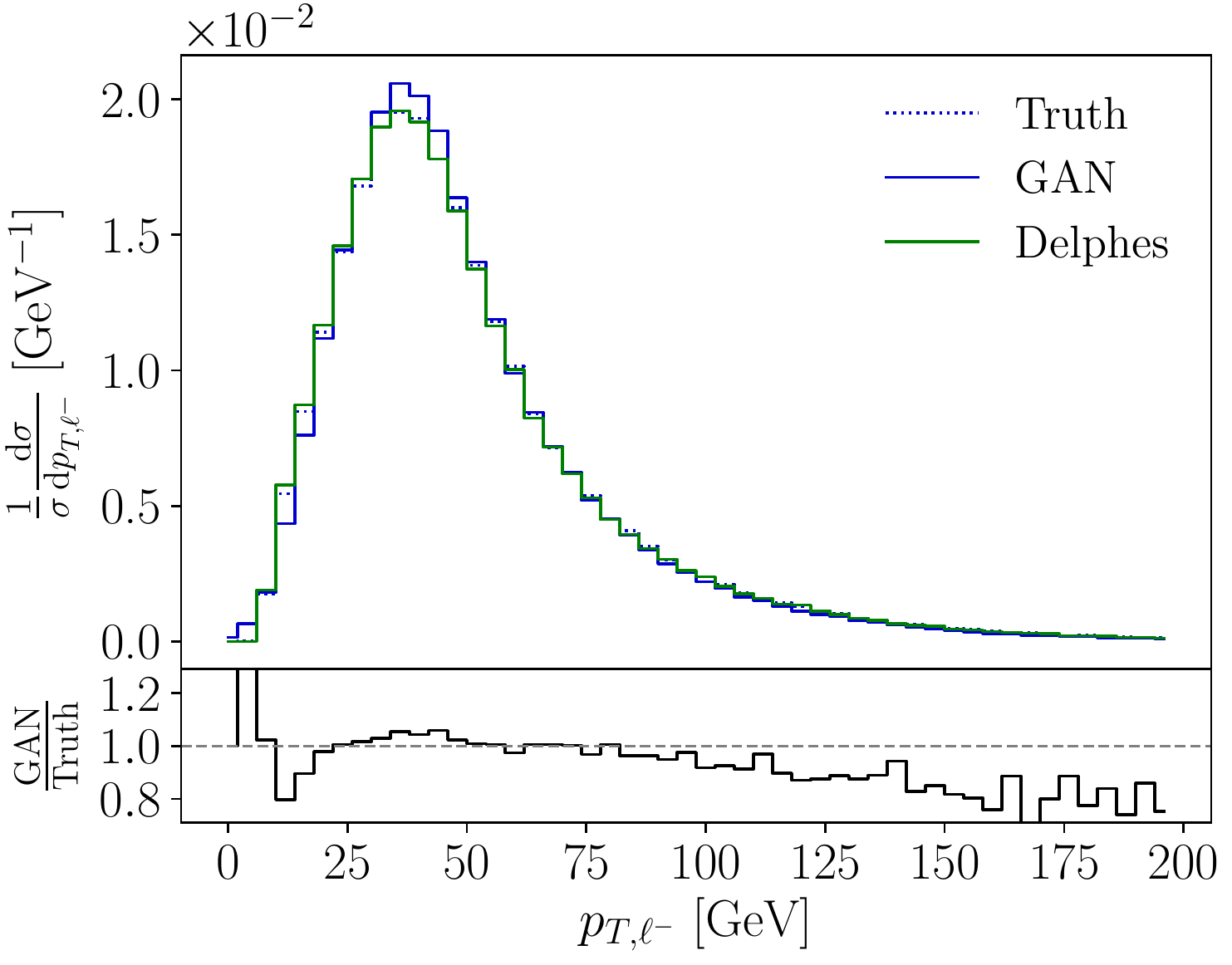}
\includegraphics[page = 4, width=0.49\textwidth]{figures/GAN_ratio} 
\caption{Example distributions for parton level truth, after
  detector simulation, and GANned back to parton level. The lower
  panels give the ratio of parton level truth and
  reconstructed parton level.}
\label{fig:distributions_GAN}
\end{figure}
For our $Z_{\ell \ell} W_{jj}$ process we generate 300k events at LO using
\madgraph~(v2.6.7) \cite{madgraph} (without any generation cuts) with the standard 
\pythia~(v8.2) shower \cite{pythia} and then simulate the detector
effects event-by-event with \delphes~(v3.3.3) \cite{delphes}  using the standard
ATLAS card. For the reconstruction of the jets we use the anti-$k_t$ jet algorithm \cite{anti_kt} with $R=0.6$ which is performed via the \fastjet~(v3.1.3) \cite{FastJet} package included in \delphes. To keep our toy setup simple we select events
with exactly two jets and a pair of same-flavor opposite-sign leptons,
specifically electrons. At the detector level both jets
are required to fulfill $p_{T, j} > 25$~GeV and $|\eta_j| <
2.5$~GeV. At detector level jets are sorted by $p_T$. We assign each
jet to a corresponding parton level object based on their angular
distance. The detector and parton level leptons are assigned based on
their charge. While the resulting smearing of the lepton momenta will
only have a modest effect, the observed widths of the hadronically
decaying $W$-boson will be much larger than the parton-level
Breit-Wigner distribution. For this reason, we focus on showing hadronic observables to benchmark the performance of our setup.

In Fig.~\ref{fig:distributions_GAN} we
compare true parton-level events to the output from a GAN trained to
unfold the detector effects.  We run the unfolding GAN on a set of
statistically independent, but simulation-wise identical sets of
detector-level events. Both, the relatively flat $p_{T,j_1}$ and the
peaked $m_{jj}$ distributions agree well between the true parton-level
events and the GAN-inverted sample, indicating that the statistical
inversion of the detector effect works well.

A great advantage of this GAN approach is that, strictly speaking,
we do not need event-by-event matched samples before and after
detector simulation. The entire training is based on batches of
typically 512 events, and these batches are independently chosen 
from the parton-level and detector-level samples. Increasing the batch size
within the range allowed by the memory size and hence reducing the impact of 
event-wise matching will actually improve the GAN training, because
it reduces statistical uncertainties~\cite{gan_phasespace}.

The big challenge arises when we want to unfold an event sample which is not
statistically equivalent to the training data; in other words, the
unfolding model is not exactly the same as the test data. As a simple
example we train the GAN on data covering the full phase space and
then apply and test the GAN on data only covering part of the
detector-level phase space. Specifically, we apply the two sets of jet
cuts
\begin{alignat}{5}
&\text{Cut I}: & \quad 
p_{T,j_1} &= 30~...~100~\gev 
\label{eq:jetcut1a} \\
&\text{Cut II}: & \quad 
p_{T,j_1} &= 30~...~60~\gev \quad \text{and} \quad p_{T,j_2} = 30~...~50~\gev \; ,
\label{eq:jetcut1b}
\end{alignat}
which leave us with 88\% and 38\% of events, respectively. This
approach ensures that the training has access to the full information,
while the test sample is a significantly reduced sub-set of the full
sample.

\begin{figure}[t]
\centering
\includegraphics[page = 1, width=0.49\textwidth]{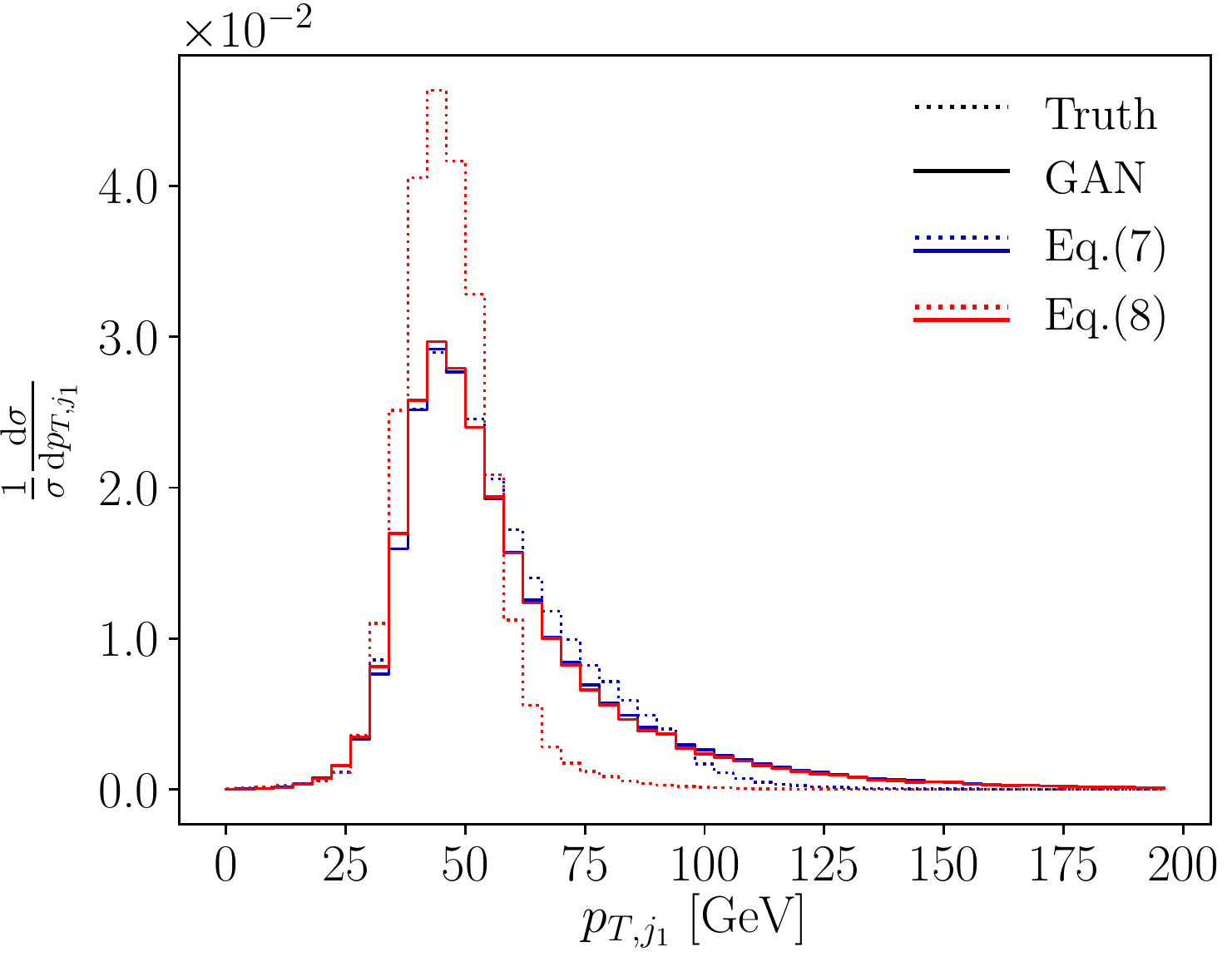}
\includegraphics[page = 2, width=0.49\textwidth]{figures/GAN_overlap} 
\caption{Parton level truth and GANned distributions when we train the
  GAN on the full data set but only unfold parts of phase space
  defined in Eq.\eqref{eq:jetcut1a} and Eq.\eqref{eq:jetcut1b}.}
\label{fig:distributions_GAN_sliced}
\end{figure}

In Fig.~\ref{fig:distributions_GAN_sliced} we show a set of kinematic
distributions, for which we GAN only part of the phase space. As
before, we can compare the original parton-level shapes of the
distributions with the results from GAN-inverting the fast detector
simulation.  We see that especially the GANned $p_{T,j}$ distribution
is strongly sculpted by the phase space cuts. This indicates that the
naive GAN approach to unfolding does not work once the training and
test data sets are not statistically identical. In a realistic
unfolding problem we cannot expect the training and test data sets to
be arbitrarily similar, so we have to go beyond the naive GAN setup
described in Fig.~\ref{fig:GANs}. The technical reason for this
behavior is that events which are similar or, by some metric, close at
the detector level are not guaranteed to be mapped onto events which
are close on the parton level. Looking at classification networks this
is the motivation to apply variational methods, for instance upgrade
autoencoders to variational autoencoders.  For a GAN we discuss a
standard solution in the next section.

\section{Fully conditional GAN}
\label{sec:fcgan}

\begin{figure}[t]
\centering
\input{incl_network_cgan}
\caption{Structure of our fully conditional FCGAN. The
  input $\{r\}$ describes a batch of random numbers and $\{ x_{G,d,p}
  \}$ denotes events sampled from the generator, detector-level data,
  or parton-level data. The blue (red) arrows indicate which
  connections are used in the training of the generator
  (discriminator).}
\label{fig:FCGAN}
\end{figure}
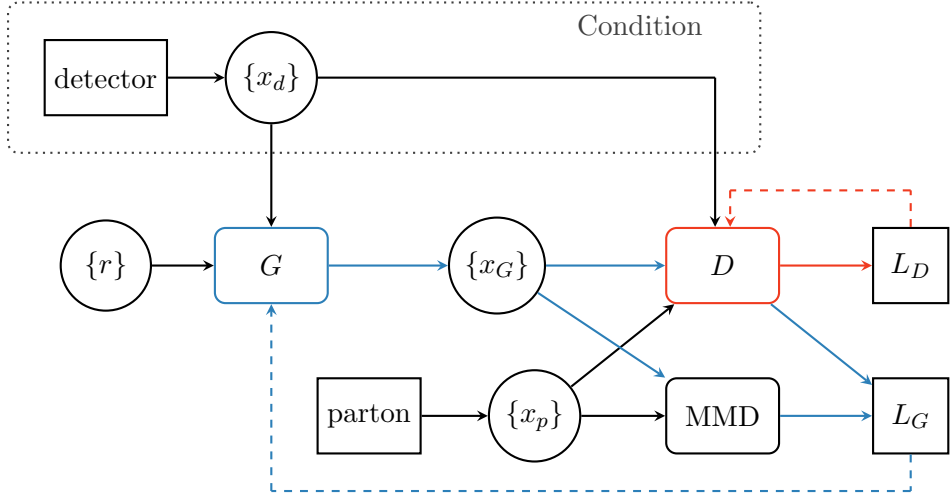

The way out of the sculpting problem when looking at different phase
space regions is to add a conditional structure to the
GAN~\cite{cond_gan} shown in Fig.~\ref{fig:GANs}. The idea behind the
conditional setup is not to learn a deterministic link between input
and output samples, because we know that without an enforced structure
in the weight or function space the generator does not benefit from
the structured input. In other words, the network does not properly
exploit the fact that the detector-level and parton-level data sets in
the training sample are paired.  A second, related problem of the
naive GAN is that once trained the model is completely deterministic,
so each detector-level event will always be mapped to the same
parton-level events. This goes against the physical intuition that
this entire mapping is statistical in nature.

\begin{figure}[t]
\centering
\includegraphics[page = 2, width=0.49\textwidth]{figures/cGAN_full_ratio}
\includegraphics[page = 3, width=0.49\textwidth]{figures/cGAN_full_ratio} \\
\includegraphics[page = 1, width=0.49\textwidth]{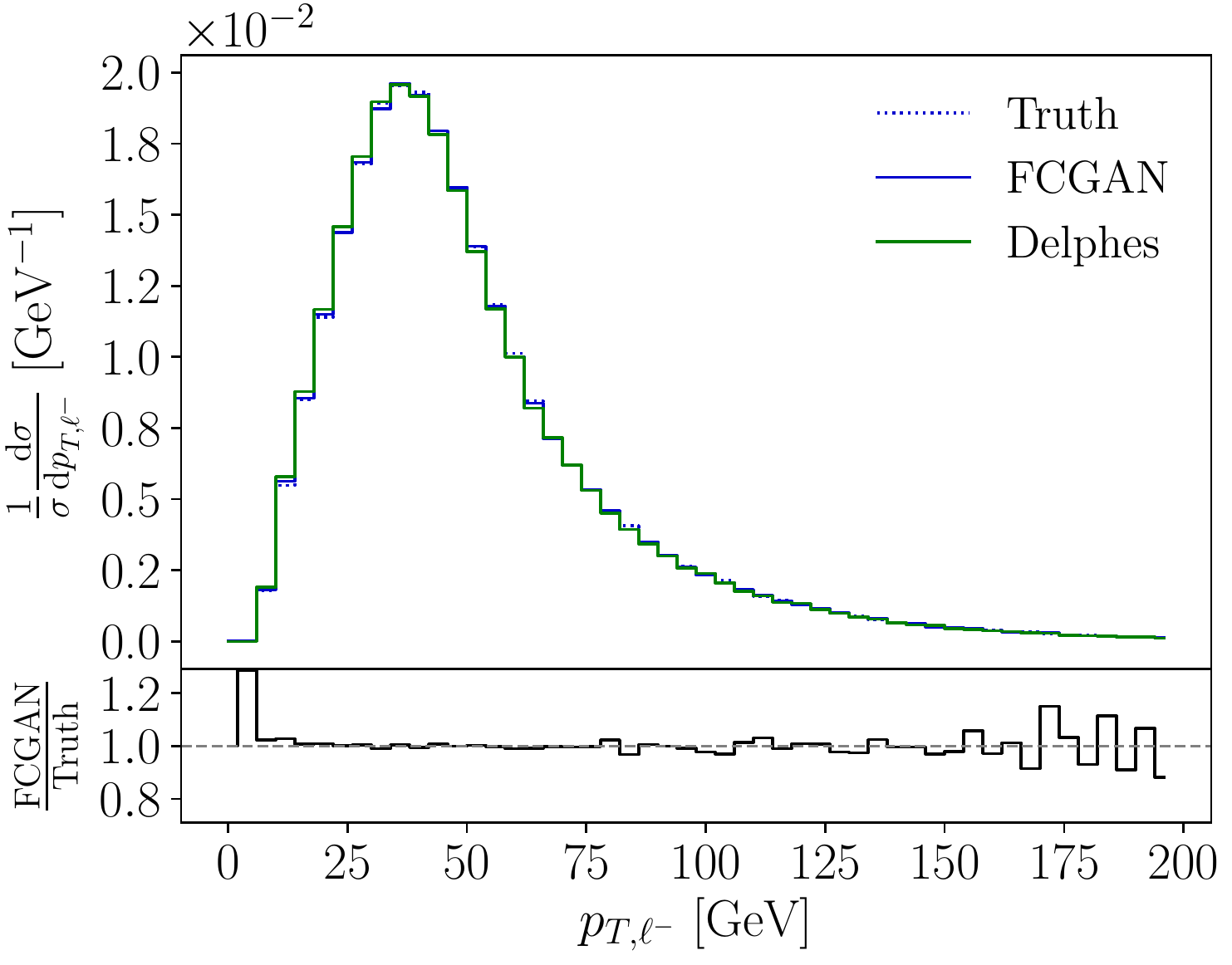}
\includegraphics[page = 4, width=0.49\textwidth]{figures/cGAN_full_ratio} 
\caption{Example distributions for parton level truth, after detector
  simulation, and FCGANned back to parton level. The lower panels give
  the ratio of parton level truth and reconstructed parton level.  The
  lower panels give the deviation between parton level truth and
  reconstructed parton level. To be compared with the naive GAN
  results in Fig.~\ref{fig:distributions_GAN}.}
\label{fig:distributions_FCGAN}
\end{figure}

\begin{table}[b!]
\begin{small} \begin{center}
\begin{tabular}{l r | l r}
\toprule
Parameter              & Value   & Parameter              & Value  \\
\midrule
Layers & 12 & Batch size & 512 \\
Units per layer & 512 & Epochs & 1200\\
Trainable weights G & 3M  & Iterations per epoch & 500\\
Trainable weights D & 3M  & Number of training events & $3 \times 10^5$\\
\midrule
$\lambda_G$ & 1 \\
$\lambda_D$ & $10^{-3}$ \\
\bottomrule
\end{tabular}
\end{center} \end{small}
\caption{FCGAN setup.}
\label{tab:details}
\end{table}

In Fig.~\ref{fig:FCGAN} we introduce a fully conditional GAN
(FCGAN). It is identical to our naive network the way we train and use
the generator and discriminator. However, the input to the generator
are actual random numbers $\{ r \}$, and the detector-level
information $\{ x_d \}$ is used as an event-by-event conditional input
on the link between a set of random numbers and the parton-level
output, \ie $G( \{ r \}, \{ x_d \} ) = \{ x_G \}$. This way the FCGAN
can generate parton-level events from random noise but still using the
detector-level information as input. To also condition the
discriminator we modify its loss to
\begin{align}
L_D \to L_D^\text{(FC)}= \left\langle - \log D\left(x, y\right) \right\rangle_{x \sim P_p, y \sim P_d} + \left\langle - \log\left( 1-D\left(x,y\right)\right) \right\rangle_{x \sim P_G, y \sim P_d} \; ,
\label{eq:D_closs}
\end{align}
and the regularized loss function changes accordingly,
\begin{align}
\begin{split}
L_D^\text{(reg)} \to L_D^\text{(reg,\,FC)} =
L_D^\text{(FC)}
&+ \lambda_D\,
\Langle \left(1- D(x,y)\right)^2 \vert \nabla \phi \vert^2 \Rangle_{x \sim P_p, y \sim P_d} \\
&+ \lambda_D\,
\Langle D\left(x,y\right)^2\, \vert \nabla \phi \vert^2 \Rangle_{x \sim P_G, y \sim P_d}  \; ,
\end{split}
\label{eq:Dcloss2}
\end{align}
again using the conventions of Ref.~\cite{gan_phasespace}. The
generator loss function now takes the form
\begin{align}
L_G\to L_G^\text{(FC)} = \left\langle - \log D\left(x,y\right) \right\rangle_{x \sim P_G, y \sim P_d} \; .
\label{eq:G_closs}
\end{align}
Note, that we do not build a conditional version of the MMD loss.  The
hyper-parameters of our FCGAN are summarized in
Tab.~\ref{tab:details}. Changing from a naive GAN to a fully
conditional GAN we have to pay a price in the structure of the
training sample. While the naive GAN only required event batches to be
matched between parton level and detector level, the training of the
FCGAN actually requires event-by-event matching.\medskip

In Fig.~\ref{fig:distributions_FCGAN} we compare the truth and the
FCGANned events, trained on and applied to events covering the full
phase space. Compared to the naive GAN, inverting the detector effects
now works even better. The systematic under-estimate of the GAN rate
in tails no longer occurs for the FCGAN.  The reconstructed invariant
$W$-mass forces the network to dynamically generate a very narrow
physical width from a comparably broad Gaussian peak. Using our usual
MMD loss developed in Ref.~\cite{gan_phasespace} we reproduce the peak
position, width, and peak shape to about 90\%. We emphasize that
the MMD loss requires us to specify the relevant one-dimensional
distribution, in this case $m_{jj}$, but it then extracts the on-shell
mass or width dynamically. The multi-kernel approach we use in this
case is explained in the Appendix.

\begin{figure}[t]
\centering
\includegraphics[page = 2, width=0.49\textwidth]{figures/cGAN_overlap_1}
\includegraphics[page = 3, width=0.49\textwidth]{figures/cGAN_overlap_1} \\
\includegraphics[page = 1, width=0.49\textwidth]{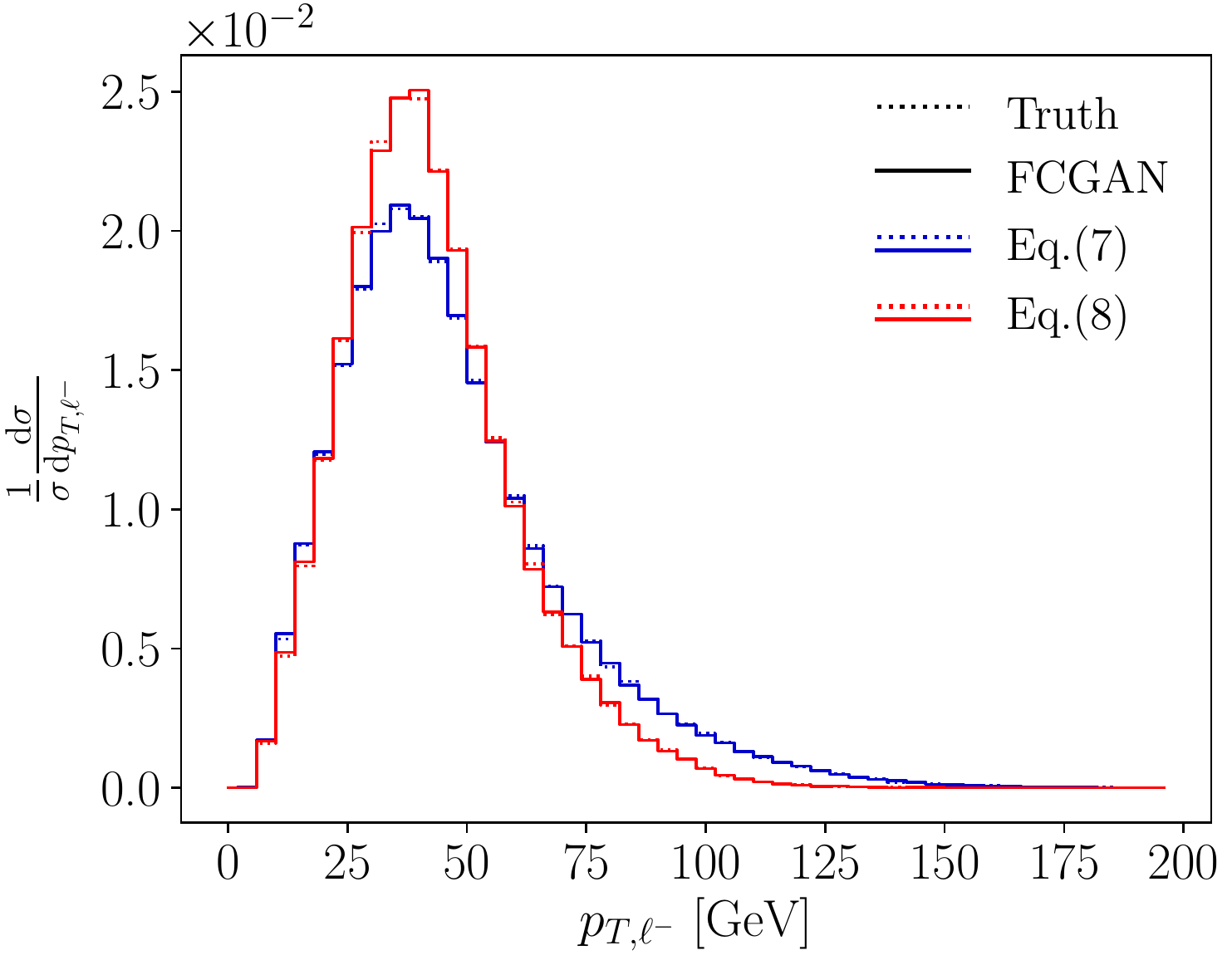}
\includegraphics[page = 4, width=0.49\textwidth]{figures/cGAN_overlap_1} 
\caption{Parton level truth and FCGANned distributions when we train
  the GAN on the full data set but only unfold parts of phase space
  defined in Eq.\eqref{eq:jetcut1a} and Eq.\eqref{eq:jetcut1b}. To be
  compared with the naive GAN results in
  Fig.\ref{fig:distributions_GAN_sliced}.}
\label{fig:distributions_FCGAN_sliced_1}
\end{figure}

As for our naive ansatz we now test what happens to the network when
the training data and the test data do not cover the same phase space
region. We train on the full set of events, to ensure that the full
phase space information is accessible to the network, but we then only
apply the network to the 88\% and 38\% of events passing the jet
cuts~I and~II defined in Eq.\eqref{eq:jetcut1a} and
Eq.\eqref{eq:jetcut1b}. We show the results in
Fig.~\ref{fig:distributions_FCGAN_sliced_1}. As observed before,
especially the jet cuts with only 40\% survival probability shape our
four example distributions. However, we see for example in the
$p_{T,jj}$ distribution that the inverted detector-level sample
reconstructs the patterns of the true parton-level events 
perfectly. This comparison indicates that the FCGAN approach deals
with differences in the training and test samples very well.

\begin{figure}[t]
\centering
\includegraphics[page = 2, width=0.49\textwidth]{figures/cGAN_overlap_2}
\includegraphics[page = 3, width=0.49\textwidth]{figures/cGAN_overlap_2} \\
\includegraphics[page = 1, width=0.49\textwidth]{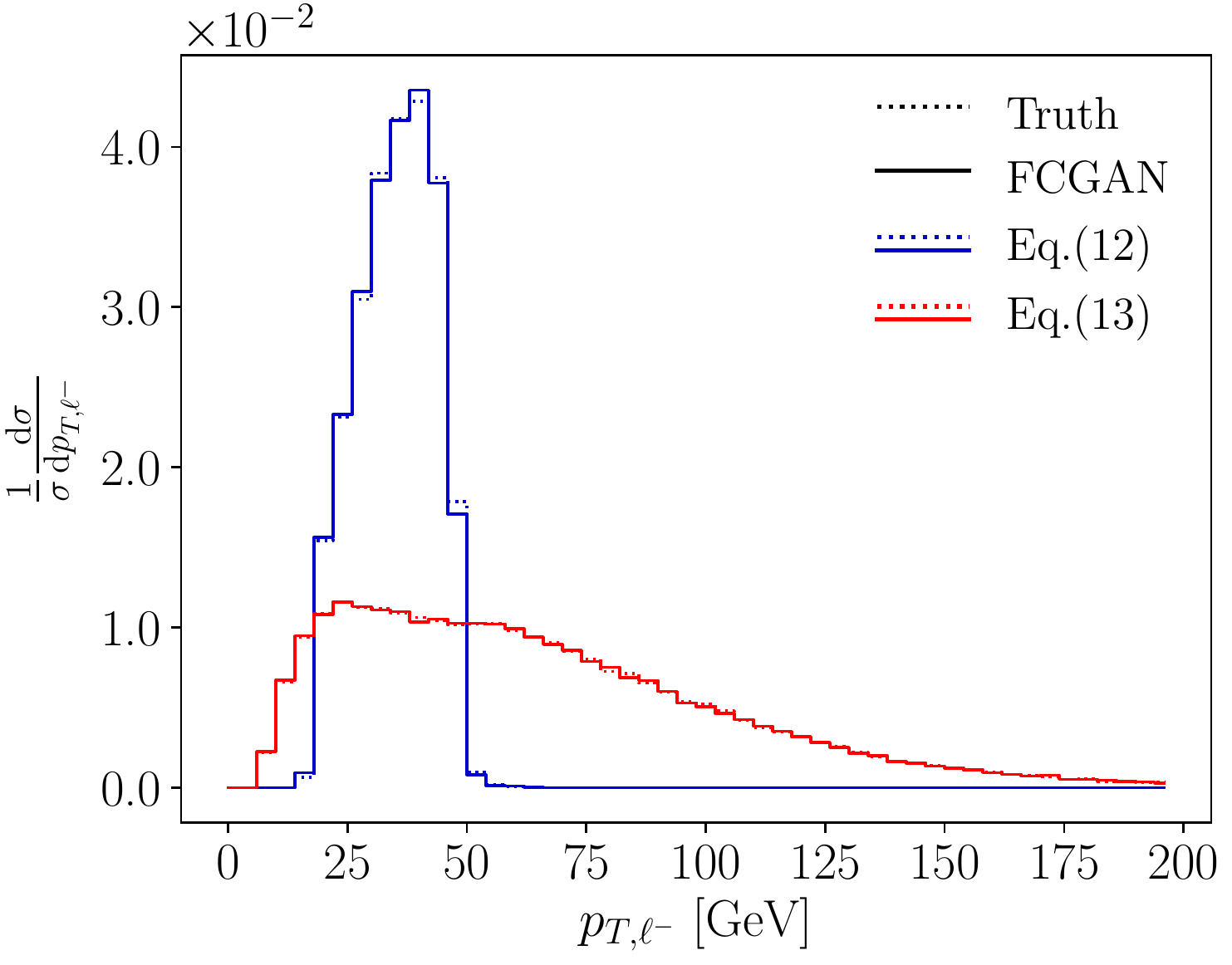}
\includegraphics[page = 4, width=0.49\textwidth]{figures/cGAN_overlap_2} 
\caption{Parton level truth and FCGANned distributions when we train
  the GAN on the full data set but only unfold parts of phase space
  defined in Eqs.\eqref{eq:jetcut2a} and~\eqref{eq:jetcut2b}.}
\label{fig:distributions_FCGAN_sliced_2}
\end{figure}

Because physicists and 4-year olds follow a deep urge to break things
we move on to harsher cuts on the inclusive event sample. We start
with
\begin{align}
\text{Cut III}: \quad p_{T,j_1}= 30~...~50~\gev
\quad p_{T,j_2} = 30~...~40~\gev
\quad p_{T,\ell^-} = 20~...~50~\gev \; ,
\label{eq:jetcut2a}
\end{align}
which 14\% of all events pass. In
Fig.~\ref{fig:distributions_FCGAN_sliced_2} we see that also for this
much reduced fraction of test events corresponding to the training
sample the FCGAN inversion reproduces the true distributions extremely
well, to a level where it appears not really relevant what fraction of
the training and test data correspond to each other.

Finally, we apply a cut which not only removes a large fraction of
events, but also cuts into the leading peak feature of the $p_{T,j_1}$
distribution and removes one of the side bands needed for an
interpolation,
\begin{align}
\text{Cut IV}: \quad  p_{T,j_1} > 60~\gev \; .
\label{eq:jetcut2b}
\end{align}
For this choice 39\% of all events pass, but we remove all events at
low transverse momentum, as can be seen from
Fig.~\ref{fig:distributions_FCGAN}. This kind of cut could therefore
be expected to break the unfolding. Indeed, the red lines in
Fig.~\ref{fig:distributions_FCGAN_sliced_2} indicate that we have
broken the $m_{jj}$ reconstruction through the FCGAN. However, all
other (shown) distributions still agree with the parton-level truth
extremely well. The problem with the invariant mass distribution is
that our implementation of the MMD loss is not actually
conditional. This can be changed in principle, but standard
implementations are somewhat inefficient and the benefit is not
obvious at this stage. At this stage it means that, when pushed
towards it limits, the network will first fail to reproduce the
correct peak width in the $m_{jj}$ distribution, while all other
kinematic variables remain stable.\medskip

\begin{figure}[t]
\centering
\includegraphics[width=0.98\textwidth]{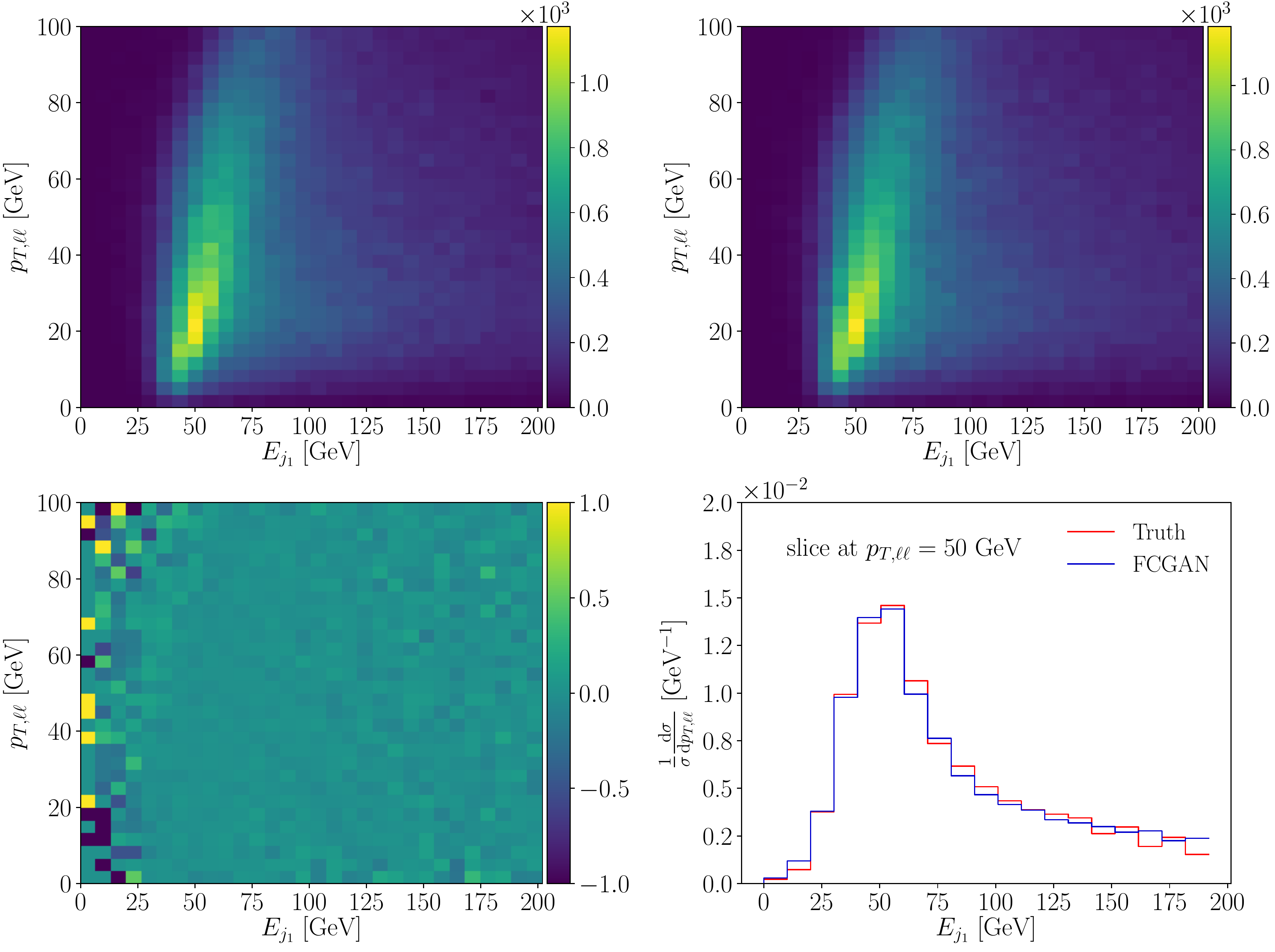}
\caption{Two-dimensional parton level truth (upper left) and FCGANned
  (upper right) distributions when we train the GAN on the full data
  set and unfold over the full phase space. The lower panels show the
  relative deviation between truth and FCGANned and the one-dimensional
  $E_{j_1}$ distribution along fixed $p_{T,\ell \ell}$.}
\label{fig:full_2d}
\end{figure}

Finally, just like in Ref.~\cite{gan_phasespace} we show 2-dimensional
correlations in Fig.~\ref{fig:full_2d}. We stick to applying the
network to the full phase space and show the parton level truth and
the FCGAN-inverted events in the two upper panels. Again, we see that
the FCGAN reproduces all features of the parton level truth with high
precision. The bin-wise relative deviation between the two 2-dimensional
distributions only becomes large for small values of $E_{j_1}$, where
the number of training events is extremely small.

\section{New physics injection}
\label{sec:closure}

As discussed before, unfolding to a hard process is necessarily
model-dependent. Until now, we have always assumed the Standard Model
to correctly describe the parton-level and detector-level events. An
obvious question is what happens if we train our FCGAN on Standard
Model data, but apply it to a different hypothesis. This challenge
becomes especially interesting if this alternative hypothesis differs
from the Standard Model in a local phase space effect. It then allows
us to test if the generator networks maps the parton-level and
detector-level phase spaces in a structured manner. Such features of
neural networks are at the heart of all variational constructions, for
instance variational autoencoders which are structurally close to
GANs. Observing them for GAN unfolding could turn into a significant
advantage over alternative unfolding methods.

To this end we add a fraction of resonant $W'$ events from a
triplet extension of the Standard Model~\cite{Biekoetter:2014jwa},
representing the hard process
\begin{align}
p p 
\to {W'}^* 
\to Z W^\pm 
\to (\ell^- \ell^+) \; (j j ) 
\end{align}
to the test data.  We simulate these events with \madgraph using the
model implementation of Ref.~\cite{Brehmer:2015rna} and denote the new
massive charged vector boson with a mass of 1.3~TeV and a width of
15~GeV as $W'$. For the test sample we combine the usual Standard
Model sample with the $W'$-sample in proportions $90\% - 10\%$.  The
other new particles do not appear in our process to leading order.
Because we want to test how well the GAN maps local phase space
structures onto each other, we deliberately choose a small width
$\Gamma_{W'}/M_{W'}\sim 1\%$, not exactly typical for such strongly
interacting triplet extensions.

\begin{figure}[t]
\centering
\includegraphics[page = 9, width=0.49\textwidth]{figures/6f_plots_mix} 
\includegraphics[page = 1, width=0.49\textwidth]{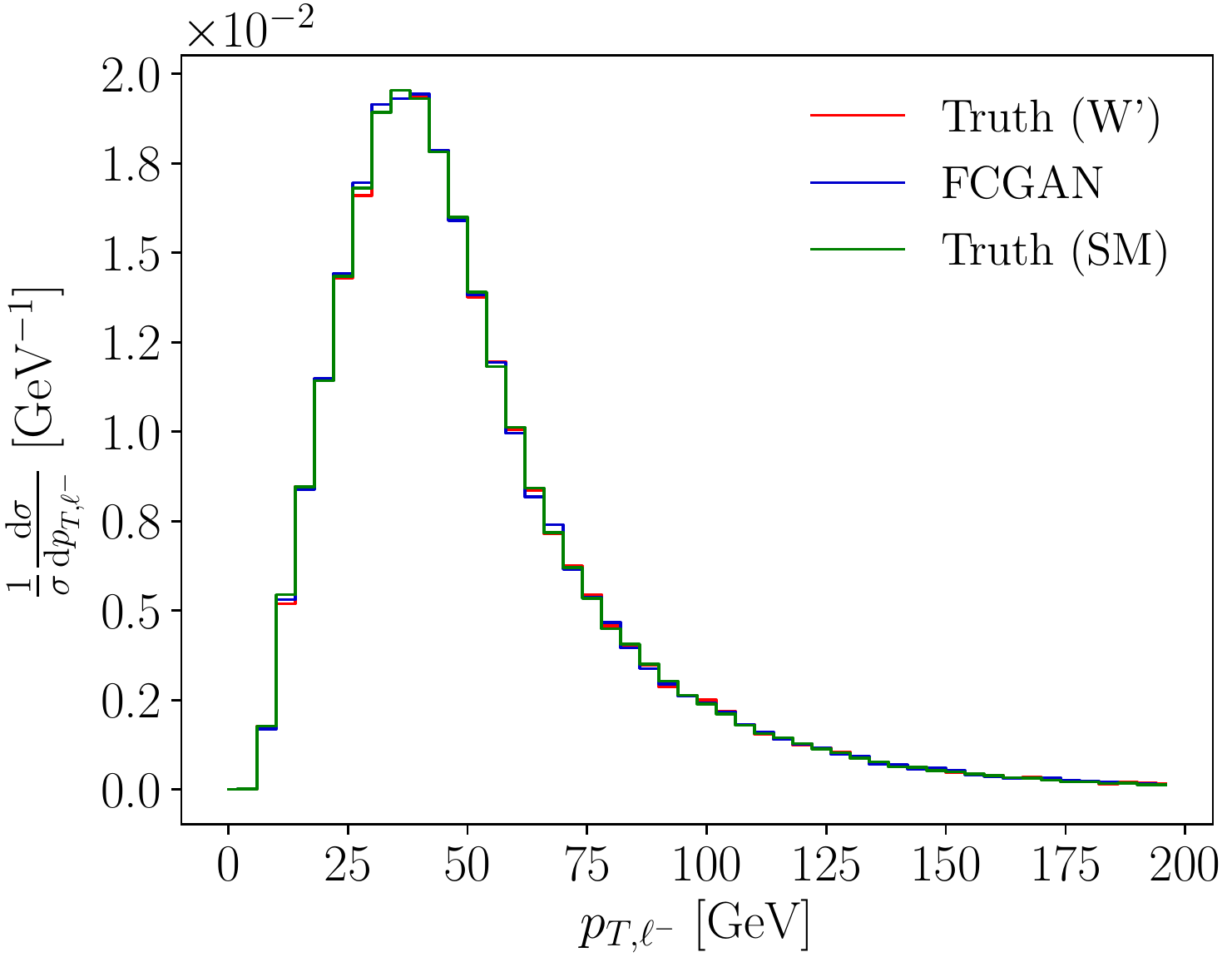}\\
\includegraphics[page = 17, width=0.49\textwidth]{figures/6f_plots_mix}
\includegraphics[page = 18, width=0.49\textwidth]{figures/6f_plots_mix} 
\caption{Parton level truth and FCGANned distributions when we train
  the network on the Standard Model only and unfold events with an
  injection of $10\%$ $W'$ events. The mass of the
  additional $s$-channel resonance is 1.3~TeV.}
\label{fig:w_prime}
\end{figure}

The results for this test are shown in Fig.~\ref{fig:w_prime}. First,
we look at transverse momentum distribution of final-state particles,
which are hardly affected by the new heavy resonance. Both, the
leading jet and the lepton distributions are essentially identical for
both truth levels and the FCGAN output. The same is true for the
invariant mass of the hadronically decaying $W$-boson, which
nevertheless provides a useful test of the stability of our training
and testing. 

Finally, we show the reconstructed $W'$-mass in the lower-right
pane. Here we see the different (normalized) truth-level distributions
for the Standard Model and the $W'$-injected sample. The FCGAN,
trained on the Standard Model, keeps track of local phase space
structures and reproduces the $W'$ peak faithfully. It also learn the
$W'$-mass as the central peak position very well. The only issue is
the $W'$-width, which the network over-estimates. However, we know
already that dynamically generated width distributions are a challenge
to GANs and require for instance an MMD loss.  Nevertheless,
Fig.~\ref{fig:w_prime} clearly shows that GAN unfolding shows a high
degree of model independence, making use of local structures in the
mapping between the two phase spaces. We emphasize that the additional
mass peak in the FCGANned events is not a one-dimensional feature, but
a localized structure in the full phase space. This local structure is
a feature of neural networks which comes in addition to the known
strengths in interpolation.

\section{Outlook}

\begin{figure}[b!]
\centering
\input{incl_flow}
\caption{Illustration of the complementary 1.FCGAN and
  \textsc{OmniFold}~\cite{omnifold} approaches.}
\label{fig:flow}
\end{figure}
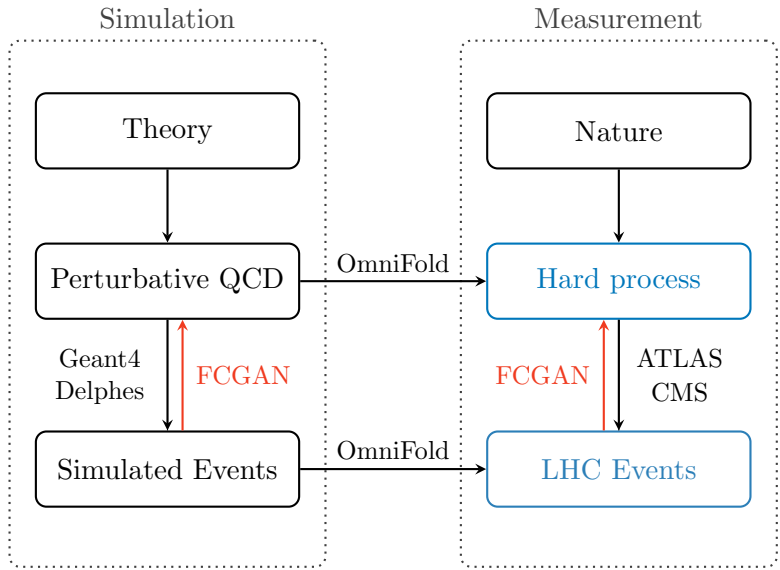

We have shown that it is possible to invert a simple Monte Carlo
simulation, like a fast detector simulation, with a fully conditional
GAN. Our example process is $WZ \to (jj) (\ell \ell)$ at the LHC and
we GAN away the effect of standard \delphes. A naive GAN approach
works extremely well when the training sample and the test sample are
very similar. In that case the GAN benefits from the fact that we do
not actually need an event-by-event matching of the parton-level and
detector-level samples.

If the training and test samples become significantly different we
need a fully conditional GAN to invert the detector effects. It maps
random noise parton-level events with conditional, event-by-event
detector-level input and learns to generate parton-level events from
detector-level events.  First, we noticed that the FCGAN with its
structured mapping provides much more stable predictions in tails of
distributions, where the training sample is statistics limited.  Then,
we have shown that a network trained on the full phase space can be
applied to much smaller parts of phase space, even including cuts in
the main kinematic features. The FCGAN successfully maintains a notion
of events close to each other at detector level and at parton level
and maps them onto each other. This approach only breaks eventually
because the MMD loss needed to map narrow Breit-Wigner propagators is
not (yet) conditional in our specific setup. 

Finally, we have seen that the network reproduces an injected new
physics signal as a local structure in phase space. This large degree
of model independence reflects another beneficial feature of neural
networks, namely the structured mapping of the linked phase spaces.

\begin{center} \textbf{1.FCGAN vs OmniFold} \end{center}

While we were finalizing our paper, the \textsc{OmniFold} approach
appeared~\cite{omnifold}. It aims at the same problem as our FCGAN,
but as illustrated in Fig.~\ref{fig:flow} it is completely
complementary. Our FCGAN uses the simulation based on \delphes to
train a generative network, which we can apply to LHC events to
generate events describing the hard process. The \textsc{OmniFold}
approach also starts from matched simulated events, but instead of
inverting the detector simulation it uses machine learning to
iteratively translate each side of this link to the measured events.
This way both approaches should be able to extract hard process
information from LHC events, assuming that we understand the relation
between perturbative QCD predictions and Monte Carlo events.

\begin{center} \textbf{Acknowledgments} \end{center}

We would like to thank Lynton Ardizzone for his extremely helpful
advice, Ben Nachman for great discussions, and Hans-Christian
Schultz-Coulon for the experimental encouragement. Concerning the
updated version, we would like to thank Ben Nachman for suggesting a
closure test like the one we are showing in Sec.~\ref{sec:closure} and
Johann Brehmer for providing the \madgraph implementation. RW and MB
acknowledge support by the IMPRS-PTFS.  The research of AB and MB is
supported by the Deutsche Forschungsgemeinschaft (DFG, German Research
Foundation) under grant 396021762 -- TRR~257 \textsl{Particle Physics
  Phenomenology after the Higgs Discovery}. GK acknowledges support by
the Deutsche Forschungsgemeinschaft under grant 390833306 -- EXC~2121
\textsl{Quantum Universe}.

\clearpage
\appendix
\section{Performance}
\label{sec:app1}

\begin{figure}[b!]
\centering
\includegraphics[page = 2, width=0.49\textwidth]{figures/pull_full}
\includegraphics[page = 3, width=0.49\textwidth]{figures/pull_full} \\
\includegraphics[page = 1, width=0.49\textwidth]{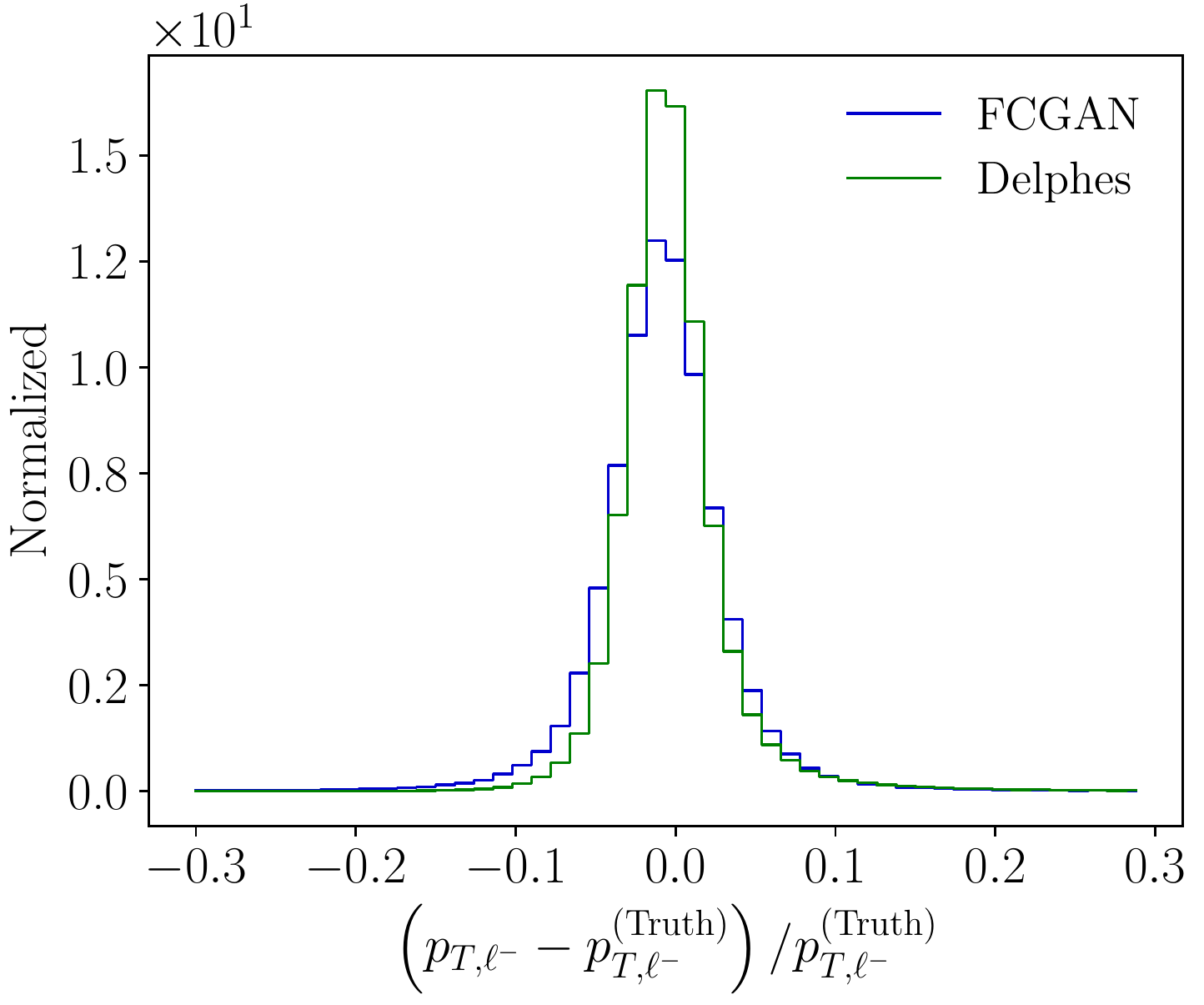}
\includegraphics[page = 4, width=0.49\textwidth]{figures/pull_full}
\caption{Normalized deviation between the FCGANned sample and truth (residual) 
 for some of the kinematic variables.}
\label{fig:app_pull}
\end{figure}

While it is clear from the main text that the FCGAN inversion of the
fast detector simulation  works extremely well, we can still show some
additional standard measures to illustrate this. For instance, in
Fig.~\ref{fig:app_pull} we show the event-wise normalized deviation
between the parton-level truth kinematics and the \delphes and
FCGAN-inverted kinematics, for instance
\begin{align}
\frac{p_{T,j}^\text{(FCGAN)} - p_{T,j}^\text{(Truth)}}{p_{T,j}^\text{(Truth)}}
\qquad \text{and} \qquad 
\frac{p_{T,j}^\text{(\delphes)} - p_{T,j}^\text{(Truth)}}{p_{T,j}^\text{(Truth)}} \; .
\end{align}
The events shown in these histograms correspond to the full phase
space inversion shown in Fig.~\ref{fig:distributions_FCGAN}, but from
the discussion in the main text it is clear that the picture does not
change when we invert only part of phase space.  As expected, we see
narrow peaks around zero, with a width in the $\pm 10\%$ range for the
jet momenta and much more narrow for the leptons, which are less
affected by detector smearing. For all distributions, but especially
the reconstructed $W$-mass, we see that the FCGAN reconstruction is
significantly closer to the parton-level truth than the \delphes
events are.

\begin{figure}[t]
\centering
\includegraphics[page = 1, width=0.49\textwidth]{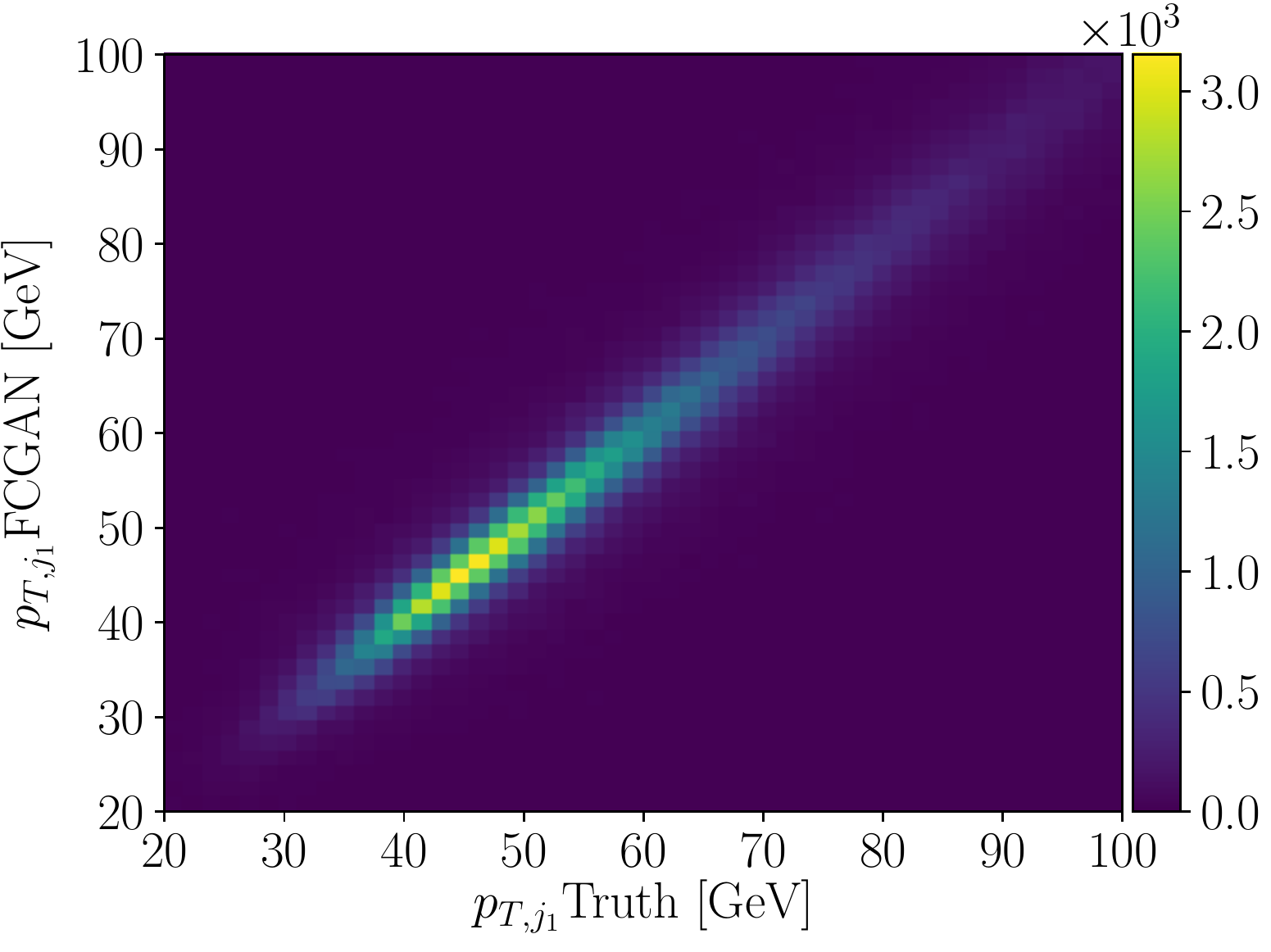}
\includegraphics[page = 1, width=0.49\textwidth]{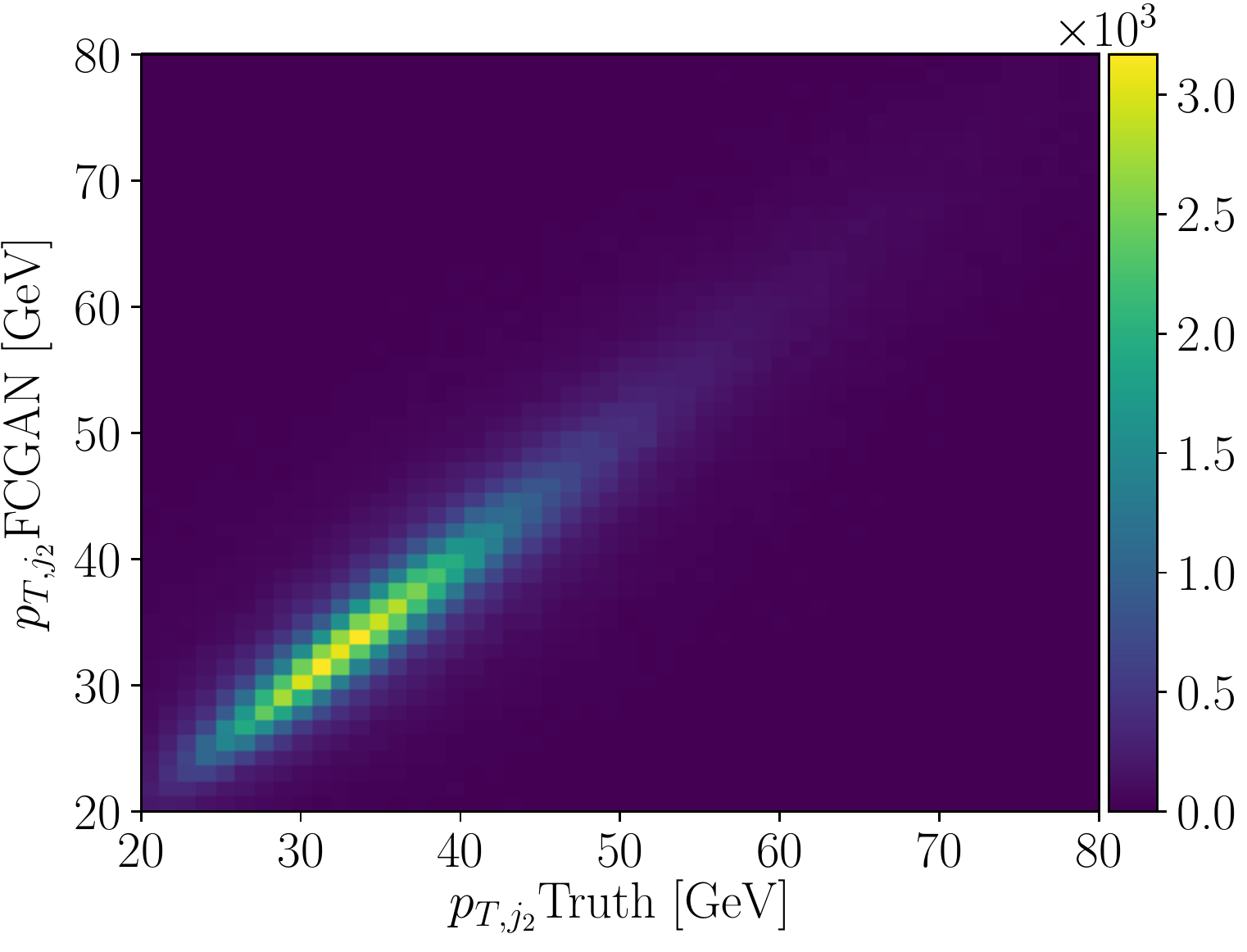}
\caption{Correlations between the FCGAN-inverted and parton-level truth 
 kinematics, or migration matrix.}
\label{fig:app_2d}
\end{figure}

Finally, we show the migration matrix or correlation between true
parton-level and reconstructed parton-level events in terms of some of
the kinematic variables in Fig.~\ref{fig:app_2d}. Not surprisingly,
we observe narrow diagonal lines.

\section{Staggered vs cooling MMD}
\label{sec:app}

The MMD loss is a two-sample test looking at the distance between
samples $x$, $x'$, drawn independently and identically distributed, in
terms of a kernel function $k\left(x, x'\right)$.  Implementations of
such a kernel, as given in Eq.~\ref{eq:kernels}, include a fixed
width or resolution $\sigma$.  We employ the MMD loss to reproduce the
invariant mass distribution of intermediate on-shell particles
$M_p$. A natural choice of $\sigma$ is the corresponding particle
width. However, this is inefficient at the beginning of the training,
when any generated invariant mass $M_G$ is essentially a random
uniform distribution. In that case $\left(x - x'\right)^2 \gg
\sigma^2$ for any $x, x' \sim M_G$, and Eq.~\ref{eq:MMD} reduces to
\begin{align}
\text{MMD}\left(k; M_G,M_P\right) \simeq \sqrt{\langle k\left(y,y'\right)\rangle_{y,y' \sim M_P}} \simeq \text{const} \; ,
\end{align}
and provides little to no gradient.

This can be avoided by computing the MMD loss using multiple kernels
with decreasing widths, so that the early training can be driven by
wide kernels.  A drawback of this approach is that only the small
subset of kernels with a resolution close to the evolving width of
$M_G$ gives a non-negligible gradient.

Alternatively, we can employ a cooling kernel, which we initialize to
some large value and then shrink to the correct particle width. This
is an efficient solution at all stages of the training.  A subtlety is
that the rate of the cooling has to follow the pace of the
generator in producing narrower invariant mass
distributions. Ultimately, we want to avoid hand-crafting the cooling
process, because it adds hyper-parameters we need to tune.  We use a
dynamic kernel width as a fixed fraction of the standard deviation of
the $M_G$ distribution.  This standard deviation as an estimate of the
width of $M_{G}$ can be replaced by any measure of the shape of
$M_{G}$, such as the full width at half maximum, and our tests show
that the performance is largely insensitive to the choice of the
fraction.

\begin{figure}[t]
\centering
\includegraphics[page=1,width=0.49\textwidth]{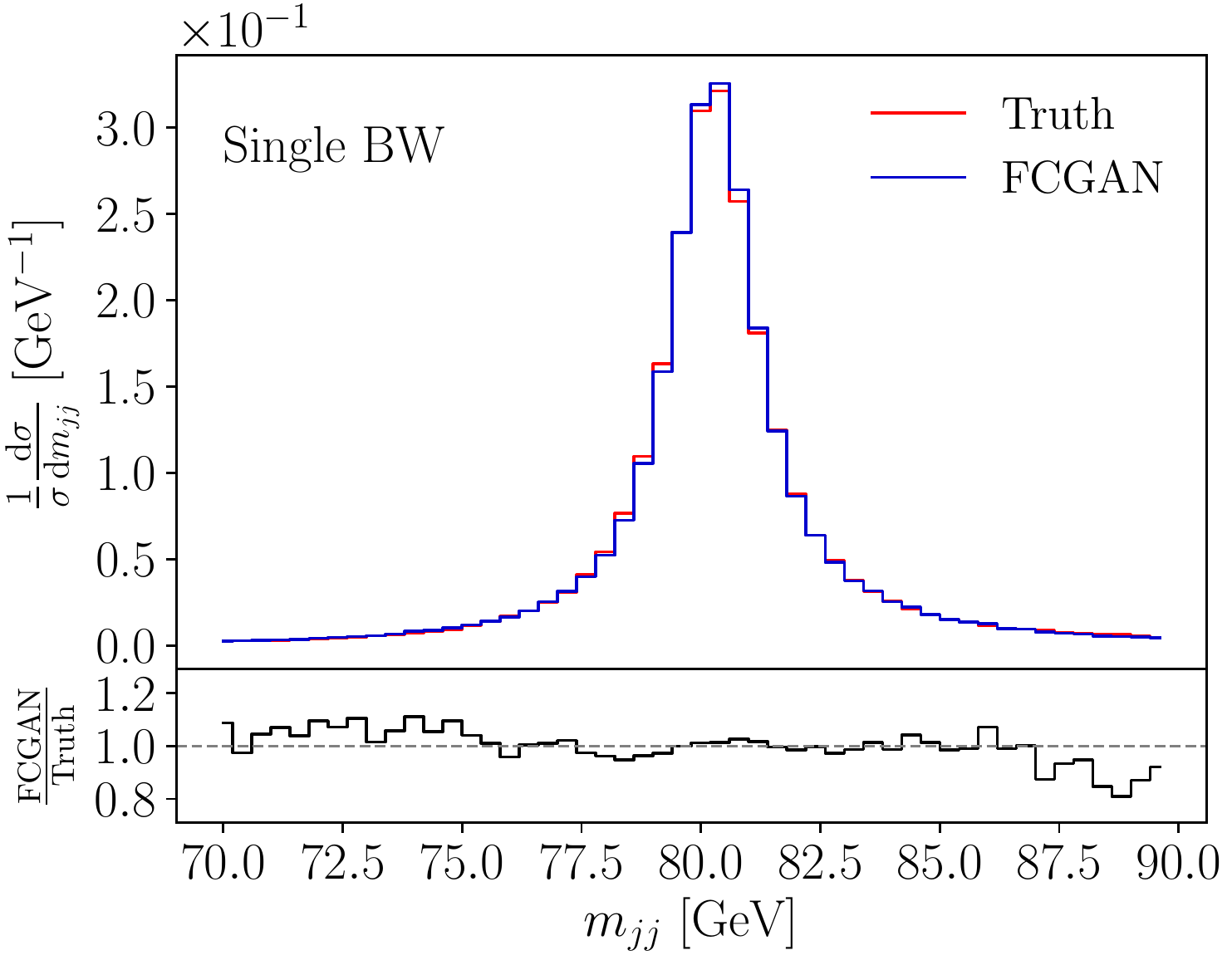}
\includegraphics[page=1,width=0.49\textwidth]{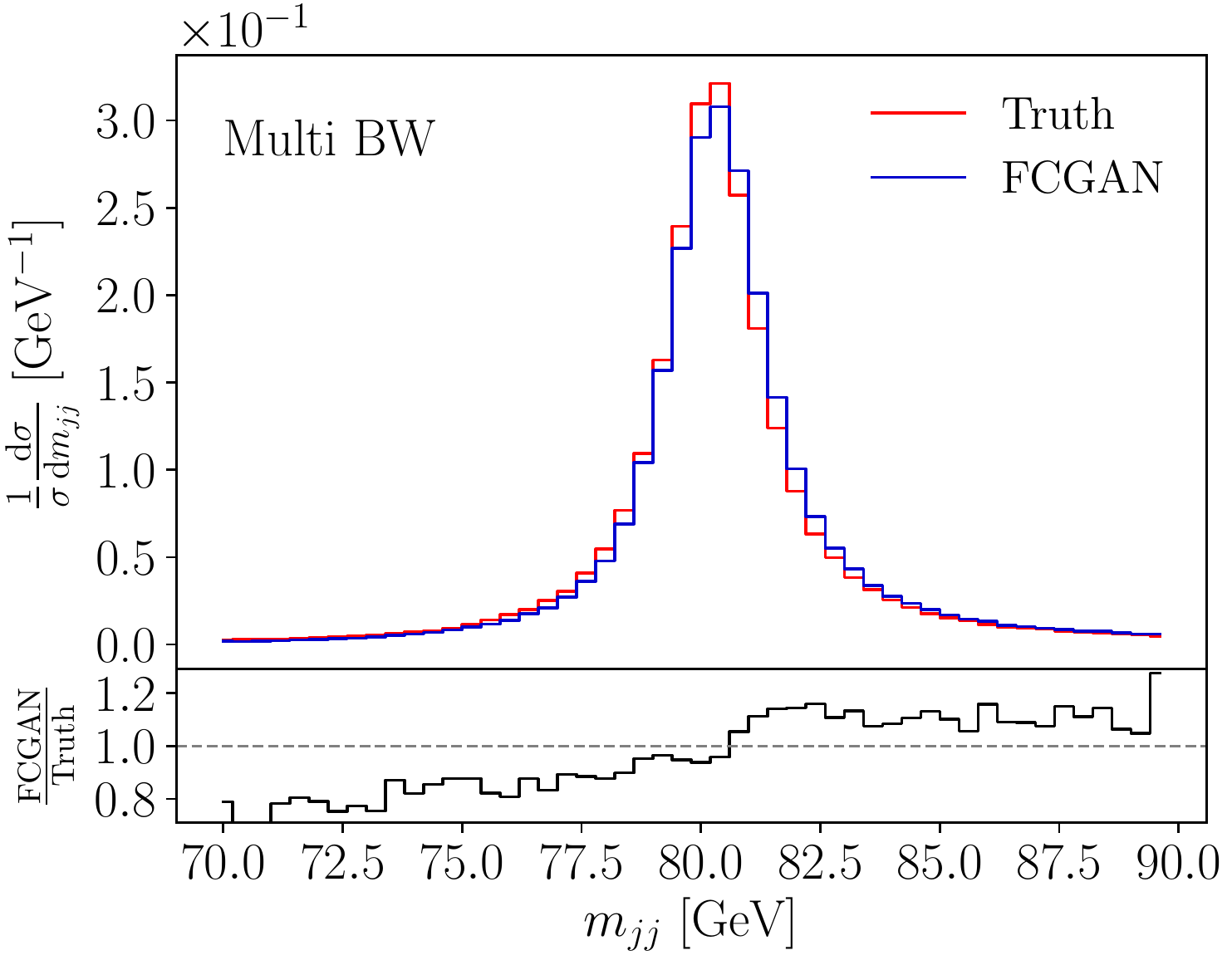}\\
\includegraphics[page=1,width=0.49\textwidth]{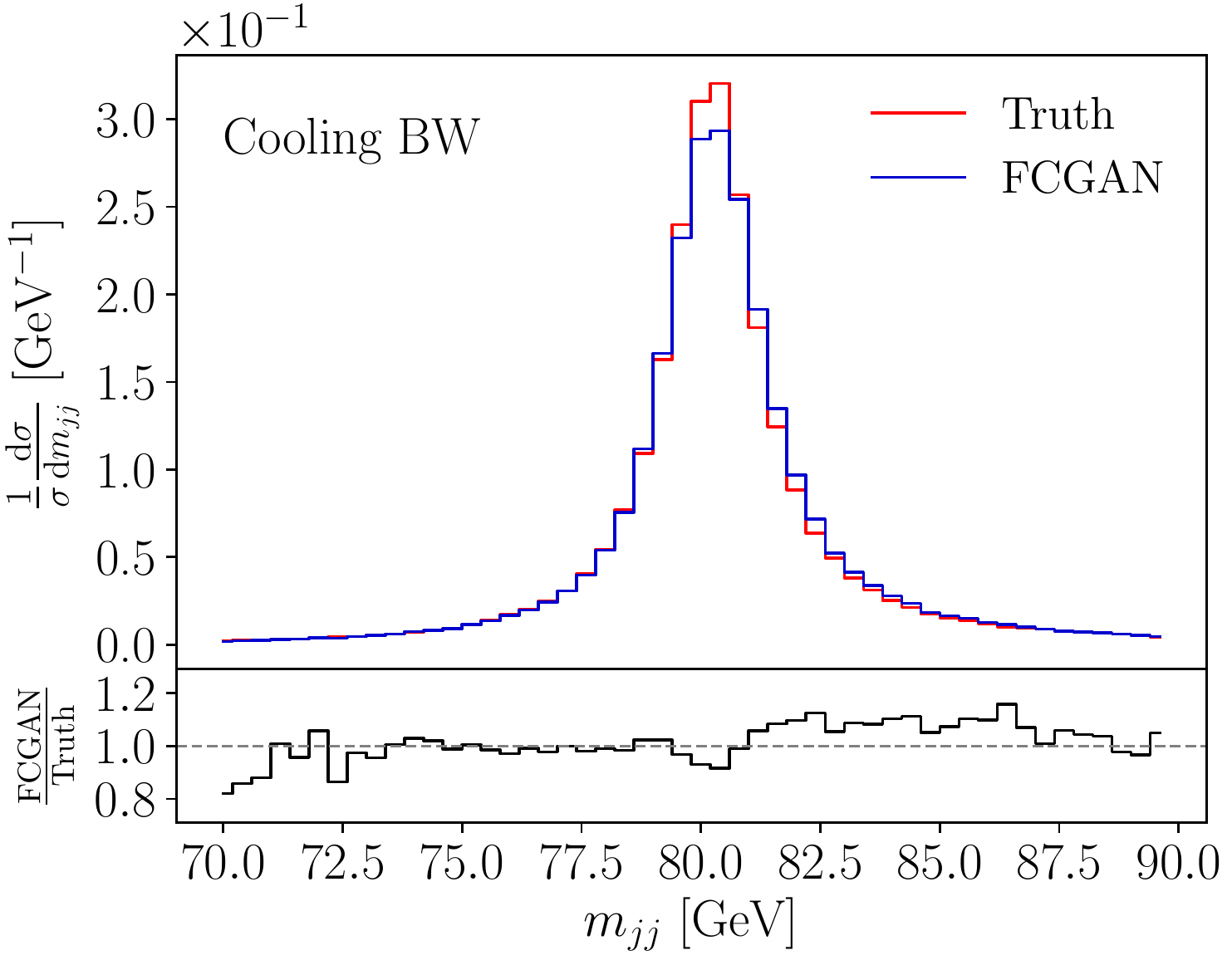}
\includegraphics[page=1,width=0.49\textwidth]{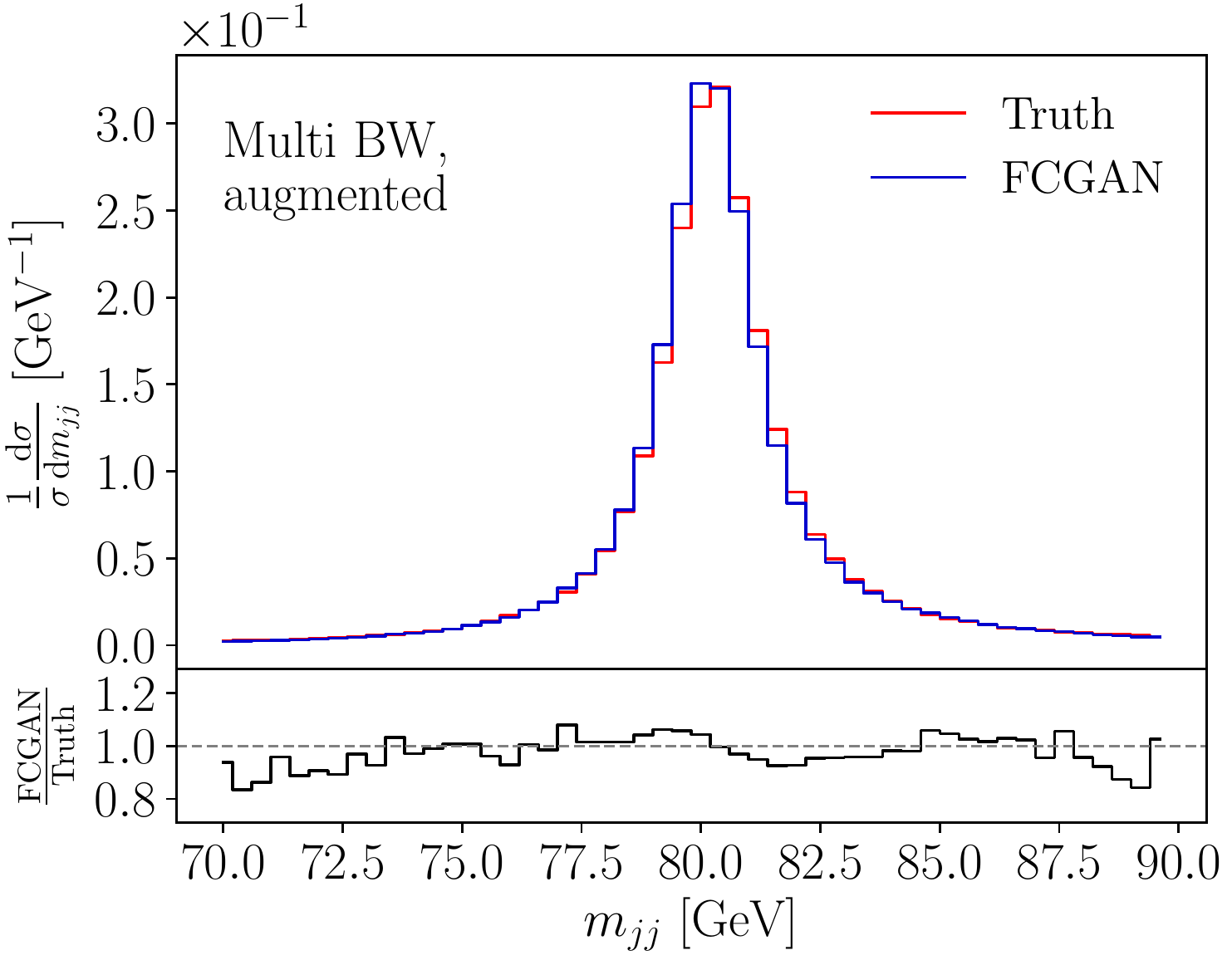}
\caption{Invariant jet-jet mass distribution for different MMD loss
  implementations: single kernel (upper left), multiple kernels (upper
  right), cooling kernel (lower left) and augmented multiple kernels
  (lower right).}
\label{fig:kernels_comparison}
\end{figure}

Yet another approach is based on the observation that the MMD kernel
test is not restricted to one-dimensional distributions
~\cite{GMMN,MomentMatching, MMDgen}.
This allows us to improve the invariant mass reconstruction by including
 additional physical information $x_i$ to the test, so that the 
 discrepancy is not computed just between the samples $M_P$ and $M_G$
 of real and generated invariant masses, but rather between 
 $(M_P, x_P)$ and $(M_G, x_G)$.
In the FCGAN spirit we therefore augment the batches of true and 
generated invariant masses with one of
conditional invariant masses. From the same detector
information used to condition the generator and the discriminator, we
can extract the detector level invariant masses $M_D$ and accordingly
compute $\text{MMD}(k; (M_G, M_D), (M_P, M_D))$. Even tough
this does not represent a conditional MMD, training with multiple
kernels benefits from using the augmented batches. In
Fig.~\ref{fig:kernels_comparison} we compare the same invariant mass
distribution using these different MMD implementations.

\clearpage
\end{fmffile}

\bibliography{literature}

\end{document}

%% file: incl_declare.tex
\definecolor{Gcolor}{HTML}{3b528b}
\definecolor{Dcolor}{HTML}{e41a1c}

\tikzstyle{generator} = [rectangle, rounded corners, minimum width=3cm, minimum height=1cm,text centered, draw=Gcolor]
\tikzstyle{discriminator} = [rectangle, rounded corners, minimum width=3cm, minimum height=1cm,text centered, draw=Dcolor]
\tikzstyle{io} = [circle, trapezium left angle=70, trapezium right angle=110, minimum width=1cm, minimum height=1cm, text centered, draw=black]

\tikzstyle{process} = [rectangle, minimum width=1cm, minimum height=1cm, text centered, draw=black]
\tikzstyle{decision} = [rectangle, minimum width=1cm, minimum height=1cm, text centered, draw=black]

\tikzstyle{arrow} = [thick,->,>=stealth]
\usepackage{xcolor}

\newcommand{\Langle}{\big\langle}
\newcommand{\Rangle}{\big\rangle}

\newcommand\one{\leavevmode\hbox{\small1\normalsize\kern-.33em1}}

\newcommand{\gev}{\text{GeV}}

\def\slashchar#1{\setbox0=\hbox{$#1$}           
   \dimen0=\wd0                                 
   \setbox1=\hbox{/} \dimen1=\wd1               
   \ifdim\dimen0>\dimen1                        
      \rlap{\hbox to \dimen0{\hfil/\hfil}}      
      #1                                        
   \else                                        
      \rlap{\hbox to \dimen1{\hfil$#1$\hfil}}   
      /                                         
   \fi}

\newcommand{\ie}{\textsl{i.e.}\;}

\newcommand{\madgraph}{\textsc{Madgraph}5\xspace}
\newcommand{\pythia}{\textsc{Pythia}8\xspace}
\newcommand{\fastjet}{\textsc{FastJet}\xspace}
\newcommand{\delphes}{\textsc{Delphes}\xspace}

\newcommand{\keras}{\textsc{Keras}\xspace}
\newcommand{\tensorflow}{\textsc{TensorFlow}\xspace}

%% file: incl_feynman.tex
\begin{fmfgraph*}(140,70)
\fmfset{arrow_len}{2mm}
\fmfset{curly_len}{3mm}
\fmfset{wiggly_len}{3mm}
\fmfstraight
\fmfleft{i1,i2}
\fmfright{o1,o2,o3,o4}
\fmf{fermion,tension=1.0,width=0.6}{i1,v1,i2}
\fmf{photon,tension=2.5,width=0.6}{v1,v2}	
\fmf{photon,tension=1.0,label=$W$,lab.side=right,width=0.6}{v2,d1}		
\fmf{photon,tension=1.0,label=$Z$,lab.side=left,width=0.6}{v2,d2}
\fmf{fermion,width=0.6}{o3,d2,o4}
\fmf{fermion,width=0.6}{o2,d1,o1}
\fmflabel{$j$}{o1}
\fmflabel{$j$}{o2}
\fmflabel{$\ell^+$}{o3}
\fmflabel{$\ell^-$}{o4}
\end{fmfgraph*}

%% file: incl_network_gan.tex


\definecolor{Gcolor}{HTML}{2c7fb8}
\definecolor{Dcolor}{HTML}{f03b20}

\tikzstyle{generator} = [thick, rectangle, rounded corners, minimum width=1.5cm, minimum height=1cm,text centered, draw=Gcolor]
\tikzstyle{discriminator} = [thick, rectangle, rounded corners, minimum width=1.5cm, minimum height=1cm,text centered, draw=Dcolor]
\tikzstyle{mmd} = [thick, rectangle, rounded corners, minimum width=1.5cm, minimum height=1cm,text centered, draw=black]
\tikzstyle{io} = [thick,circle, trapezium left angle=70, trapezium right angle=110, minimum width=1.2cm, minimum height=1cm, text centered, draw=black]

\tikzstyle{iodotted} = [thick, circle, trapezium left angle=70, trapezium right angle=110, minimum width=1.2cm, minimum height=1cm, text centered, draw=black, dotted]

\tikzstyle{process} = [thick, rectangle, minimum width=1cm, minimum height=1cm, text centered, draw=black]

\tikzstyle{xG} = [thick,rectangle, minimum width=2.2cm, minimum height=3cm, text depth= 2.2cm, draw=black]
\tikzstyle{s0} = [thick,rectangle, minimum width=2cm, minimum height=3cm, text centered]
\tikzstyle{s1} = [thick, dotted, rectangle, minimum width=1.6cm, minimum height=1.1cm, text centered, draw=black]

\tikzstyle{decision} = [thick,rectangle, minimum width=1cm, minimum height=1cm, text centered, draw=black]

\tikzstyle{dots} = [circle, minimum size=2pt, inner sep=0pt,outer sep=0pt, draw=Dcolor, fill = Dcolor]

\tikzstyle{arrow} = [thick,->,>=stealth]

\begin{tikzpicture}[node distance=2cm]

\node (generator) [generator] {$G$};
\node (xG) [io, right of = generator, xshift=1.0cm, yshift=0cm] {$\{x_G\}$};
\node (xd) [io, left of=generator, xshift=-0.2cm, yshift=0cm] {$\{x_d\}$};
\node (discriminator) [discriminator, right of = xG, xshift=1.0cm, yshift=0cm] {$D$};

\node (xp) [io, below of = xG, xshift=0.5cm, yshift=0cm] {$\{x_p\}$};

\node (detector) [process, below of=xd, xshift=-0.0cm, yshift=0cm] {detector};
\node (parton) [process, left of=xp, xshift=-0.2cm, yshift=0cm] {parton};

\draw [arrow, color=black] (detector) -- (xd);
\draw [arrow, color=black] (parton) -- (xp);
\draw [arrow, color=black] (xd) -- (generator);
\draw [arrow, color=black] (xp) -- (discriminator);
\draw [arrow, color=Gcolor] (generator) -- (xG);
\draw [arrow, color=Gcolor] (xG) -- (discriminator);

\node (dloss) [process, right of=discriminator, xshift=0.5cm, yshift=0cm] {$L_{D}$};
\node (gloss) [process, below of=dloss, xshift=0.0cm, yshift=0cm] {$L_{G}$};
\node (mmd) [mmd, below of=discriminator, xshift=0.0cm, yshift=0cm] {MMD};

\draw [arrow, color=Gcolor] (discriminator) -- (gloss);
\draw [arrow, color=Dcolor] (discriminator) -- (dloss);

\coordinate[ above of = dloss, xshift=0cm, yshift=-1cm] (d1);
\coordinate[ above of = discriminator, xshift=0cm, yshift=-1cm] (d2);
\draw[thick, dashed, color=Dcolor] (dloss) -- (d1);
\draw[thick, dashed, color=Dcolor] (d1) -- (d2);
\draw[arrow, dashed, color=Dcolor] (d2) -- (discriminator);

\draw[arrow, color=Gcolor] (xG) --  (mmd);
\draw[arrow, color=black] (xp) --  (mmd);
\draw[arrow, color=Gcolor] (mmd) --  (gloss);

\coordinate[ below of = gloss, xshift=0cm, yshift=1.0cm] (out1);
\coordinate[ below of = generator, xshift=0cm, yshift=-1.0cm] (out2);
\draw[thick, dashed, color=Gcolor] (gloss) --  (out1);
\draw[thick, dashed, color=Gcolor] (out1) --  (out2);
\draw[arrow, dashed, color=Gcolor] (out2) --  (generator);

\end{tikzpicture}

%% file: incl_network_cgan.tex

\definecolor{Gcolor}{HTML}{2c7fb8}
\definecolor{Dcolor}{HTML}{f03b20}

\tikzstyle{generator} = [thick, rectangle, rounded corners, minimum width=1.5cm, minimum height=1cm,text centered, draw=Gcolor]
\tikzstyle{discriminator} = [thick, rectangle, rounded corners, minimum width=1.5cm, minimum height=1cm,text centered, draw=Dcolor]
\tikzstyle{mmd} = [thick, rectangle, rounded corners, minimum width=1.5cm, minimum height=1cm,text centered, draw=black]
\tikzstyle{io} = [thick,circle, trapezium left angle=70, trapezium right angle=110, minimum width=1.2cm, minimum height=1cm, text centered, draw=black]

\tikzstyle{cond} = [thick, rectangle, dotted, rounded corners, minimum width=10.0cm, minimum height=2cm,text centered, draw=gray!50!black]

\tikzstyle{iodotted} = [thick, circle, trapezium left angle=70, trapezium right angle=110, minimum width=1.2cm, minimum height=1cm, text centered, draw=black, dotted]

\tikzstyle{process} = [thick, rectangle, minimum width=1cm, minimum height=1cm, text centered, draw=black]

\tikzstyle{xG} = [thick,rectangle, minimum width=2.2cm, minimum height=3cm, text depth= 2.2cm, draw=black]
\tikzstyle{s0} = [thick,rectangle, minimum width=2cm, minimum height=3cm, text centered]
\tikzstyle{s1} = [thick, dotted, rectangle, minimum width=1.6cm, minimum height=1.1cm, text centered, draw=black]

\tikzstyle{decision} = [thick,rectangle, minimum width=1cm, minimum height=1cm, text centered, draw=black]

\tikzstyle{dots} = [circle, minimum size=2pt, inner sep=0pt,outer sep=0pt, draw=Dcolor, fill = Dcolor]

\tikzstyle{arrow} = [thick,->,>=stealth]

\begin{tikzpicture}[node distance=2cm]

\node (generator) [generator] {$G$};
\node (random) [io, left of=generator, xshift=-0.2cm, yshift=0cm] {$\{ r \}$};
\draw [arrow, color=black] (random) -- (generator);
\node (xG) [io, right of = generator, xshift=1.0cm, yshift=0cm] {$\{x_G\}$};
\node (discriminator) [discriminator, right of = xG, xshift=1.0cm, yshift=0cm] {$D$};

\node (cond) [cond, above of = generator, xshift=1.5cm, yshift=0.5cm] {};
\node (condi) [above of = xG, xshift=1.9cm, yshift=1.2cm, color=gray!50!black] {Condition};

\node (xd) [io, above of = generator, xshift=0.cm, yshift=0.5cm] {$\{x_d\}$};
\node (xp) [io, below of = xG, xshift=0.5cm, yshift=0cm] {$\{x_p\}$};

\node (detector) [process, left of=xd, xshift=-0.2cm, yshift=0cm] {detector};
\node (parton) [process, left of=xp, xshift=-0.2cm, yshift=0cm] {parton};

\coordinate[ above of= discriminator, xshift=-0.1cm, yshift=0.5cm] (in1);
\draw [thick, color=black] (xd) -- (in1);
\draw [arrow, color=black] (in1) -- ([xshift=-0.1cm] discriminator.north);

\draw [arrow, color=black] (detector) -- (xd);
\draw [arrow, color=black] (parton) -- (xp);
\draw [arrow, color=black] (xd) -- (generator);
\draw [arrow, color=black] (xp) -- (discriminator);
\draw [arrow, color=Gcolor] (generator) -- (xG);
\draw [arrow, color=Gcolor] (xG) -- (discriminator);

\node (dloss) [process, right of=discriminator, xshift=0.5cm, yshift=0cm] {$L_{D}$};
\node (gloss) [process, below of=dloss, xshift=0.0cm, yshift=0cm] {$L_{G}$};
\node (mmd) [mmd, below of=discriminator, xshift=0.0cm, yshift=0cm] {MMD};

\draw [arrow, color=Gcolor] (discriminator) -- (gloss);
\draw [arrow, color=Dcolor] (discriminator) -- (dloss);

\coordinate[ above of = dloss, xshift=0cm, yshift=-1cm] (d1);
\coordinate[ above of = discriminator, xshift=0.1cm, yshift=-1cm] (d2);
\draw[thick, dashed, color=Dcolor] (dloss) -- (d1);
\draw[thick, dashed, color=Dcolor] (d1) -- (d2);
\draw[arrow, dashed, color=Dcolor] (d2) -- ([xshift=0.1cm] discriminator.north);

\draw[arrow, color=Gcolor] (xG) --  (mmd);
\draw[arrow, color=black] (xp) --  (mmd);
\draw[arrow, color=Gcolor] (mmd) --  (gloss);

\coordinate[ below of = gloss, xshift=0cm, yshift=1.0cm] (out1);
\coordinate[ below of = generator, xshift=0cm, yshift=-1.0cm] (out2);
\draw[thick, dashed, color=Gcolor] (gloss) --  (out1);
\draw[thick, dashed, color=Gcolor] (out1) --  (out2);
\draw[arrow, dashed, color=Gcolor] (out2) --  (generator);

\end{tikzpicture}

%% file: incl_flow.tex


\definecolor{Gcolor}{HTML}{f03b20}
\definecolor{pcolor}{HTML}{0077bb}
\definecolor{dcolor}{HTML}{2c7fb8}


\tikzstyle{theory} = [thick, rectangle, rounded corners, minimum width=1.5cm, minimum height=1cm,text centered, draw=Gcolor]
\tikzstyle{nature} = [thick, rectangle, rounded corners, minimum width=1.5cm, minimum height=1cm,text centered, draw=pcolor]
\tikzstyle{none} = [thick, rectangle, rounded corners, minimum width=3.5cm, minimum height=1cm,text centered, draw=black]
\tikzstyle{io} = [thick,circle, trapezium left angle=70, trapezium right angle=110, minimum width=1.2cm, minimum height=1cm, text centered, draw=black]

\tikzstyle{cond} = [thick, rectangle, dotted, rounded corners, minimum width=4.2cm, minimum height=7cm,text centered, draw=gray!50!black]

\tikzstyle{iodotted} = [thick, circle, trapezium left angle=70, trapezium right angle=110, minimum width=1.2cm, minimum height=1cm, text centered, draw=black, dotted]

\tikzstyle{process} = [thick, rectangle, minimum width=1cm, minimum height=1cm, text centered, draw=black]

\tikzstyle{xG} = [thick,rectangle, minimum width=2.2cm, minimum height=3cm, text depth= 2.2cm, draw=black]
\tikzstyle{s0} = [thick,rectangle, minimum width=2cm, minimum height=3cm, text centered]
\tikzstyle{s1} = [thick, dotted, rectangle, minimum width=1.6cm, minimum height=1.1cm, text centered, draw=black]

\tikzstyle{decision} = [thick,rectangle, minimum width=1cm, minimum height=1cm, text centered, draw=black]

\tikzstyle{dots} = [circle, minimum size=2pt, inner sep=0pt,outer sep=0pt, draw=Dcolor, fill = Dcolor]

\tikzstyle{arrow} = [thick,->,>=stealth]

\begin{tikzpicture}[node distance=2cm]

\node (theory) [none] {Theory};
\node (nature) [none, right of = theory, xshift = 4cm] {Nature};

\node (parton) [none, below of = theory, color = black] {Perturbative QCD};
\node (hard) [none, color = pcolor, below of = nature] {Hard process};

\node (mcevent) [none, below of = parton, yshift=-0.5cm, color = black] {Simulated Events};
\node (data) [none, color = dcolor, below of = hard, yshift=-0.5cm] {LHC Events};

\node (theorybig)[cond, below of = theory, yshift=-0.3cm]{};
\node (theorybig)[cond, below of = nature, yshift=-0.3cm]{};
\node(theo1)[above of = theory, color=gray!50!black,yshift=-0.5cm]{Simulation};
\node(theo1)[above of = nature, color=gray!50!black,yshift=-0.5cm]{Measurement};

\draw[arrow] (theory) -- (parton);
\draw[arrow] (nature) -- (hard);
\draw[arrow] (parton) --  node[scale=0.9, below, anchor=center, xshift=-1.0cm, yshift=0.25cm] {Geant4} (mcevent);
\draw (parton) -- node[scale=0.9, below, anchor=center, xshift=-1.0cm, yshift=-0.25cm] {Delphes} (mcevent);

\draw[arrow] (hard) --node[scale=0.9, above, anchor=center, xshift=0.9cm, yshift=0.25cm] {ATLAS}  (data);
\draw (hard) --node[scale=0.9, above, anchor=center, xshift=0.9cm, yshift=-0.25cm] {CMS}  (data);

\draw[arrow] (parton) -- node [scale=0.9, above] {OmniFold} (hard);
\draw[arrow] (mcevent) -- node [scale=0.9, above] {OmniFold}  (data);

\draw[arrow, color=Gcolor] ([xshift=0.2cm] mcevent.north) -- node [scale=0.9,below, anchor=center, xshift=0.9cm ] {FCGAN} ([xshift=0.2cm]parton.south);
\draw[arrow, color=Gcolor] ([xshift=-0.2cm] data.north) -- node [scale=0.9,above,anchor=center, xshift=-0.9cm ] {FCGAN} ([xshift=-0.2cm]hard.south);

\end{tikzpicture}